\documentclass[fleqn,usenatbib]{mnras}
\usepackage{newtxtext,newtxmath}

\usepackage[T1]{fontenc}

\usepackage{amsmath}
\usepackage{multirow}
\usepackage[dvips]{graphicx}
\usepackage[caption=false]{subfig}
\usepackage{hyperref}
\usepackage{cleveref}
\usepackage{threeparttable}
\usepackage[dvipsnames]{xcolor}
\usepackage{ulem}

\graphicspath{{./}{figures/}}

\newcommand{\nodata}{\dots}

\usepackage{slashbox}

\title[Polarization variability in AGN jets]{MOJAVE XXI. Decade-long linear polarization variability in AGN jets at parsec scales}

\author[Zobnina et al.]{D.~I.~Zobnina,$^{1}$\thanks{E-mail: di.zobnina@gmail.com}
H.~D.~Aller,$^{2}$
M.~F.~Aller,$^{2}$
D.~C.~Homan,$^{3}$
Y.~Y.~Kovalev,$^{4,1,5}$
M.~L.~Lister,$^{6}$
\newauthor
I.~N.~Pashchenko,$^{1}$
A.~B.~Pushkarev,$^{7,1}$
T.~Savolainen$^{8,9,4}$\\
$^{1}$Astro Space Centre of Lebedev Physical Institute, Profsoyuznaya 84/32, Moscow 117997, Russia\\
$^{2}$Department of Astronomy, University of Michigan, Ann Arbor, MI 48109-1107 USA\\
$^{3}$Department of Physics and Astronomy, Denison University, Granville, OH 43023, USA\\
$^{4}$Max-Planck-Institut f\"ur Radioastronomie, Auf dem H\"ugel 69, D-53121 Bonn, Germany\\
$^{5}$Moscow Institute of Physics and Technology, Institutsky per. 9, Dolgoprudny, Moscow region, 141700, Russia\\
$^{6}$Department of Physics and Astronomy, Purdue University, 525 Northwestern Avenue, West Lafayette, IN 47907, USA\\
$^{7}$Crimean Astrophysical Observatory, 298409 Nauchny, Crimea, Russia\\
$^{8}$Aalto University Department of Electronics and Nanoengineering, PL 15500, FI-00076 Aalto, Finland\\
$^{9}$Aalto University Mets\"ahovi Radio Observatory, Mets\"ahovintie 114, FI-02540 Kylm\"al\"a, Finland
}

\date{Accepted 2023 May 11. Received 2023 May 11; in original form 2022 July 29}
\pubyear{2023}

\begin{document}
\label{firstpage}
\pagerange{\pageref{firstpage}--\pageref{lastpage}}
\maketitle

\begin{abstract}
Using stacking of images obtained at different epochs, we studied the variability properties of linear polarization of active galactic nucleus (AGN) jets on parsec-scales. Our sample is drawn from the MOJAVE programme, and consists of 436 AGNs manifesting core-jet morphology and having at least five VLBA observing epochs at 15~GHz from January 1996 through August 2019, with some additional archival VLBA data reduced by us. We employed a stacking procedure and constructed maps of (i) standard deviation of fractional polarization and electric vector position angle (EVPA) over epochs as the measure of variability and (ii) median polarization degree to quantify typical values in time. The distributions of these values along and across the jet were analysed for the whole sample for the first time.  We found that core EVPA variability is typically higher than that of the jet, presumably due to component blending and outflow bends in the core. The BL~Lacertae object cores have lower EVPA variability, compared to that of quasars, possibly due to lower Faraday rotation measure, suggesting a stronger ordered magnetic field component. The EVPA becomes more stable down the jet. Most of the sources showing this trend have a time coverage of more than 12~years and at least 15~epochs. The possible cause could be the increase of stability in the magnetic field direction, reflecting an increase in the fraction of the magnetic field that is ordered. There are no significant optical-class-dependent or spectral-class-dependent relations in the EVPA variability properties in AGN jets.

\end{abstract}

\begin{keywords}
quasars: general -- BL~Lacertae objects: general --  galaxies: active -- galaxies: jets -- radio continuum: galaxies -- polarization
\end{keywords}

\section{Introduction} \label{sec:intro}

A large-scale magnetic field plays a key role in launching relativistic jets \citep{Blandford_1977, Blandford_1982, Lovelace_1987}, and its toroidal component effectively collimates the jets \citep{Benford_1978, Chan_1980}. The magnetic field manifests itself through linearly polarized synchrotron emission and the Faraday rotation effect.
	
In BL~Lac objects (BL~Lacs), the electric vector position angle (EVPA) often aligns with the local jet direction \citep{Gabuzda_2000, Lister_2005, O'Sullivan_2009}, while quasars show a distribution with no preferred direction \citep{Lister_2005}. The analysis of first-epoch maps of 133 extragalactic jets (mainly quasars and BL~Lacs) obtained within the MOJAVE VLBA programme at 15~GHz showed that the typical fractional polarization of AGN jet components varies from 3~per~cent for quasars to 10~per~cent for BL~Lacs \citep[][]{Lister_2005}. The gap between quasars and BL~Lacs is large, with a possible reason being the more common presence of strong, transverse shocks \citep[][and  references therein]{Lister_2005, Aller_2017} or a stronger helical field component in BL~Lacs. \citet{Wardle_2013} considered a combination of uniform and isotropic random magnetic field. In this scenario, the difference in fractional polarization could be explained by a difference in viewing angles provided that the quasar viewing angle is on average larger than the BL~Lac one.
\cite{MOJAVE_XXI} studied time-averaged polarization properties of AGN jets using the same observational data, as in this paper. 
Their analysis showed comparable fractional polarization of quasar and BL~Lac jets starting from hecto-parsec de-projected scales and beyond.
Also, they confirmed the tendency of the EVPAs to align with the local jet direction for BL~Lacs; for quasars, the EVPAs are predominantly orthogonal to the ridgeline. They found a significant rise of the typical fractional polarization downstream in the jet and toward its edges. The latter could be a manifestation of a helical magnetic field or a shear layer.

AGNs are known to be extremely variable in total intensity and polarized emission in the radio band, with the time-scales of variations ranging from less than one day to several years. The intra-day variability can be either intrinsic \citep[e.g.][]{Gabuzda_2000a, Gabuzda_2000b, Kravchenko_2020} or produced by scintillation in the ionized interstellar medium \citep{Rickett_1984} or gravitational microlensing \citep{Nottale_1986}. An example of intrinsic variations is the recent Event Horizon Telescope observations of 3C~279 with $\sim$20~$\mu$as angular resolution at 230~GHz \citep{Kim_2020}, which reveal significant closure phase day-to-day changes, which seem to be connected to two components non-radially propagating down the inner jet. Some observations show the correlation of EVPA rotations on the time-scales of several days with a new superluminal knot emerging from the core \citep[e.g. a review paper by][]{Park_2022}.

Our study considers long-period linear polarization variability with a median time coverage of about seven years. Such variations may be associated with a bright feature moving along the magnetic field \citep{Marscher_2008} or along bent trajectories \citep{Gomez_1994}. \citet{Cohen_2018} analysed $\sim$180$^{\circ}$ EVPA rotations of the blazar OJ~287 and interpreted them using a steady polarized jet model with two successive outbursts with counter-rotating EVPAs. Also, variability could be related to geometric effects such as Doppler boosting changes due to jet precession, magneto-hydrodynamic shocks propagating through the jet or they could reflect spatial and temporal evolution of the magnetic field. The main goals of our study are to investigate the statistical properties of linear polarization variability over the parsec-scale AGN jets to quantify the magnitude and understand the reason for the variations.
 
\citet{Hodge_2018} studied linear polarization variability in the core region using MOJAVE data at 15~GHz. The core at 15~GHz is a partially optically thick region near the jet base. One of their results is that the core components exhibiting the highest fractional polarization and lowest variability in fractional polarization also have stable EVPAs that align closely with the local jet direction. The authors suggested a standing transverse shock as an explanation of such magnetic field topology and its stability. Also, they found that the EVPAs of BL~Lac cores are less variable compared to those of flat spectrum radio quasars (FSRQs) and are often well-aligned with the local jet direction in contrast to FSRQs which have a tendency for misalignment. The authors suggested to explain this observation by inherent differences in shock strength and geometry of the emission region between the two optical classes.
	
This paper is comprised of the following parts: the observational data and our sample are described in Section~\ref{sec:sample}; the procedure used for the construction of variability maps is presented in Section~\ref{sec:maps}; the results of the analysis are given and discussed in Section~\ref{sec:results}; Section~\ref{sec:sum} is a summary; the detailed description of our uncertainty estimation approach is presented in Appendix~\ref{appendix:errors}; the procedure of beam full width at half-maximum (FWHM) size calculation is given in Appendix~\ref{appendix:beam_fits}; Appendix~\ref{appendix:AGNs} presents notes on AGNs which show atypical behaviour of median fractional polarization, EVPA or fractional polarization variability along the jet ridgeline (Section~\ref{subsec:var_along}).

\section{Our sample} \label{sec:sample}

We formed a sample of AGNs using observational data from the MOJAVE VLBA programme\footnote{\url{https://www.cv.nrao.edu/MOJAVE}}. The sample is biased in favour of the AGNs with bright compact radio emission on milliarcsecond scales and is described in detail in \citet{Lister_2018}. For our polarization variability study, we selected 438 AGNs which have at least five VLBA epochs between 1996 January 19 and 2019 August 4. This is the minimum number of epochs for which, on the one hand, the estimator of variability as a standard deviation is evaluated accurately and, on the other hand, the number of AGNs with a lower number of observations is small (7~per~cent of the full sample). One of the sources, quasar 2023+335, was found to be affected by strong refractive-dominated scattering \citep{Pushkarev_2013}, which alters the observed position and the properties of the core. Therefore, we excluded 2023+335 from the sample. Also, the epochs where the total intensity ($I$) image had an rms noise exceeding three times the median rms noise (as determined from the images at all available epochs of a source) were removed because these noisy maps increase the rms of the total intensity images averaged over the epochs. For 1329$-$126, this requirement left only four epochs; hence, it was excluded from the sample too. Finally, we have 436 sources with 5846 maps and 367 individual epochs. Mostly, these are quasars (59~per~cent, 259 sources) and BL~Lac objects (31~per~cent, 136 sources), also 6~per~cent (24 sources) are radio~galaxies. The rest are narrow-line Seyfert~1 galaxies and optically unidentified sources (\autoref{t:opt_sed_class}). The classes of the synchrotron peak position in the spectral energy distribution (SED) are low-synchrotron-peaked (LSP) if the rest-frame peak is at a frequency below $10^{14}$~Hz, intermediate-synchrotron-peaked (ISP) if the peak is between $10^{14}$~Hz and $10^{15}$~Hz and high-synchrotron-peaked (HSP) if the peak is above $10^{15}$~Hz. The sample is strongly dominated by LSP sources because one of the selection criteria for an object to be included in the pool of MOJAVE targets was the VLBA flux density at 15~GHz to exceed 1.5~Jy. HSP AGNs tend to be weak in the radio band since their SED is skewed toward higher frequencies. The synchrotron peak classification of our sample is given in \autoref{t:opt_sed_class}. For BL~Lacs with unknown redshift $z$, the synchrotron peak position class was estimated using the median $z$ of AGNs of this optical class from the sample (0.27). These BL~Lacs are shown in a separate row in \autoref{t:opt_sed_class}. In the case of sources with unidentified optical class, the median redshift of the sample AGNs (0.79) was taken. The vast majority of AGNs have apparent one-sided jet morphology due to Doppler boosting effects; nine sources (0128+554, 0238$-$084, 0305+039, 0316+413, 1228+126, 1413+135, 1509+054, 1957+405, 2043+749) exhibit counter-jet features on milliarcsecond scales.

\begin{table}
\caption{Optical and synchrotron peak classifications.
\label{t:opt_sed_class}
}
\centering
\begin{tabular}{lccccc}
\hline\hline\noalign{\smallskip}
 & LSP & ISP & HSP & Unknown & Total \\
\hline
Quasars & 251 & 5 & \nodata & 3 & 259 \\
BL~Lacs with known $z$ & 41 & 11 & 22 & \nodata & 74 \\
BL~Lacs with unknown $z$ & 34 & 24 & 4 & \nodata & 62 \\
Radio~Galaxies & 20 & 3 & \nodata & 1 & 24 \\
Narrow-line Seyfert~1 Gal. & 5 & 1 & \nodata & \nodata & 6 \\
Unidentified & 8 & 1 & \nodata & 2 & 11 \\
Total & 359 & 45 & 26 & 6 & 436 \\
\hline
\end{tabular}
\end{table}

\begin{table*}
\centering
\caption{General characteristics of the AGN sample.
\label{t:sample_properties}
}
\begin{threeparttable}
\begin{tabular}{llcccl}
\hline\hline\noalign{\smallskip}
Source & Alias &  Opt. Class & SED Peak Class & $z$ & $z$ and/or Opt. Class Reference \\
(1) & (2) & (3) & (4) & (5) & (6) \\
\hline
0003$-$066 & NRAO 005 & B & LSP & 0.347 & \citet{2005PASA...22..277J} \\ 
0003+380 & S4 0003+38 & Q & LSP & 0.229 & \citet{1994AAS..103..349S} \\ 
0006+061 & TXS 0006+061 & B & LSP & \nodata & \citet{2012AA...538A..26R} \\ 
0007+106 & III Zw 2 & G & LSP & 0.089 & \citet{1970ApJ...160..405S} \\ 
0010+405 & 4C +40.01 & Q & LSP & 0.256 & \citet{1992ApJS...81....1T} \\ 
\hline
\end{tabular}
\begin{tablenotes}
\item
Columns are as follows:
(1) Source name;
(2) Alias;
(3) Optical Class: quasar (Q), BL~Lac (B), radio~galaxy (G), narrow-line Seyfert~1 galaxy (N), unidentified (U);
(4) Class of Synchrotron Peak Position in SED;
(5) Redshift;
(6) Redshift and/or Optical Class Reference.
This table is available in its entirety in a machine-readable form in the Supplementary data online and at the CDS VizieR. The first five entries are shown here for guidance.
\end{tablenotes}
\end{threeparttable}
\end{table*}

\begin{figure}
    \centering
    \includegraphics[width=\linewidth]{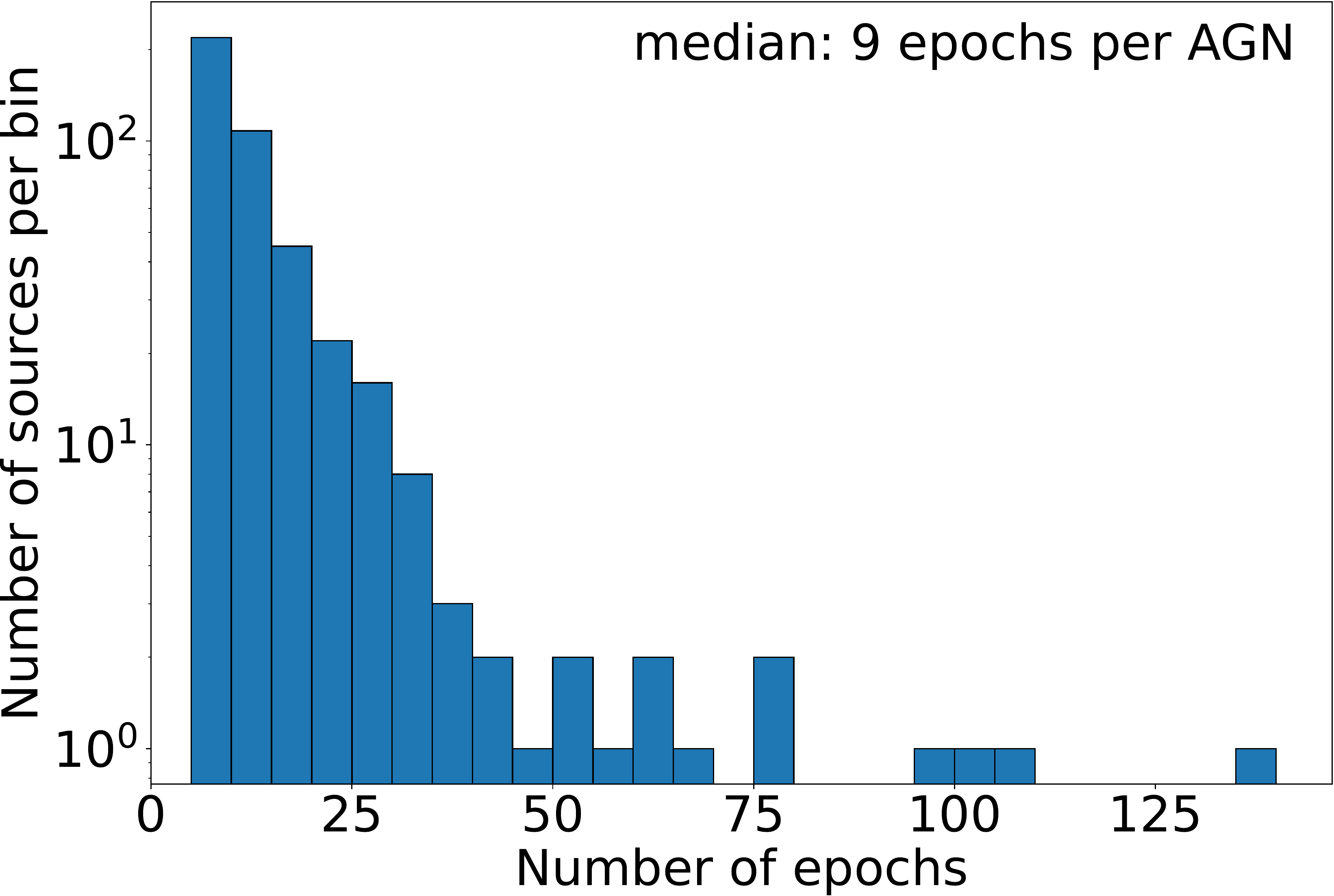}
    \caption{Histogram of the number of VLBA epochs per AGN.
    \label{f:epoch_number}
    }
\end{figure}

\begin{figure}
    \centering
    \includegraphics[width=\linewidth]{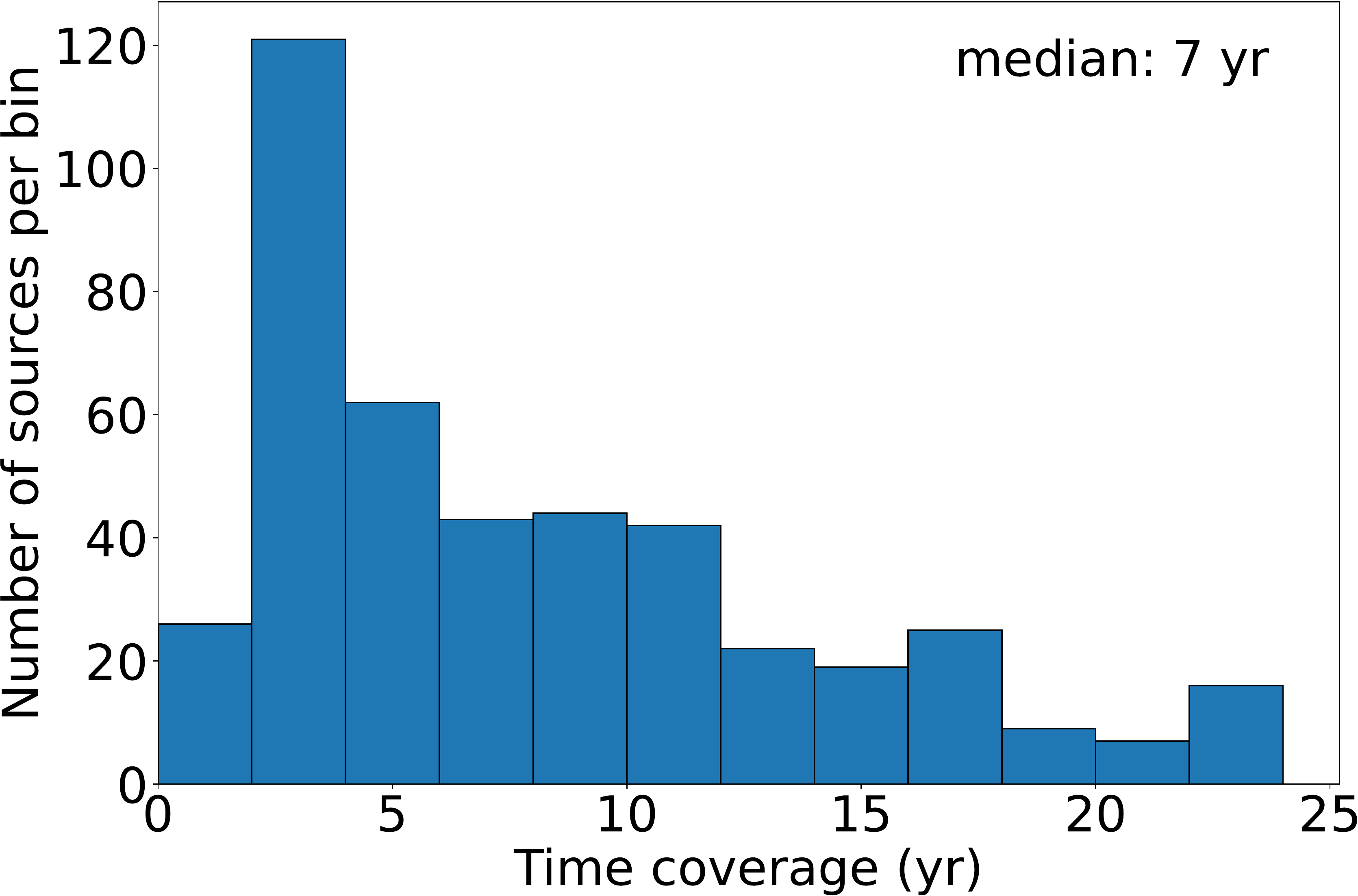}
    \caption{Histogram of the distribution of the time periods between the first and the last VLBA epochs for all sources in the sample.
    \label{f:time_coverage}
    }
\end{figure}

\autoref{t:sample_properties} provides general information on the sample sources. The distribution of the number of epochs per source is wide, ranging from 5 to 139, with a median number of 9 (Fig.~\ref{f:epoch_number}).
The time period between the first and the last epoch of observations of each individual source is distributed from one year to 24~years (Fig.~\ref{f:time_coverage}). There are two main sub-samples here: (i) one consisting of $\gamma$-ray-bright sources with high-confidence associations with sources detected by the \textit{Fermi}-LAT after 2008 \citep{Acero_2015}, which were observed with a narrower time coverage, and (ii) AGNs monitored on longer time-scales, from the archival VLBA observations 
and the MOJAVE programme. We aim to probe the full jet width to study the polarization variations not only along but also across the jet. The jet width revealed by stacking depends on the time coverage. The longer a source is observed, the wider the jet appears on the map until eventually the entire jet cross-section is filled. \citet{Pushkarev_2017} found that a time coverage $\gtrsim 5$~years typically allows revealing the full jet cross-section in total intensity at 15~GHz. About 60~per~cent of our sample meets this condition. We assume that the full width of a jet structure is seen on the total intensity maps averaged over epochs; however, the increase in the jet cross-section with time coverage could be explained by instabilities developing down the jet or by the oscillatory behaviour of the inner jet orientation seen in some frequently observed sources \citep{Lister_2013,Lister_2021}.

\begin{figure}
    \centering
    \includegraphics[width=\linewidth]{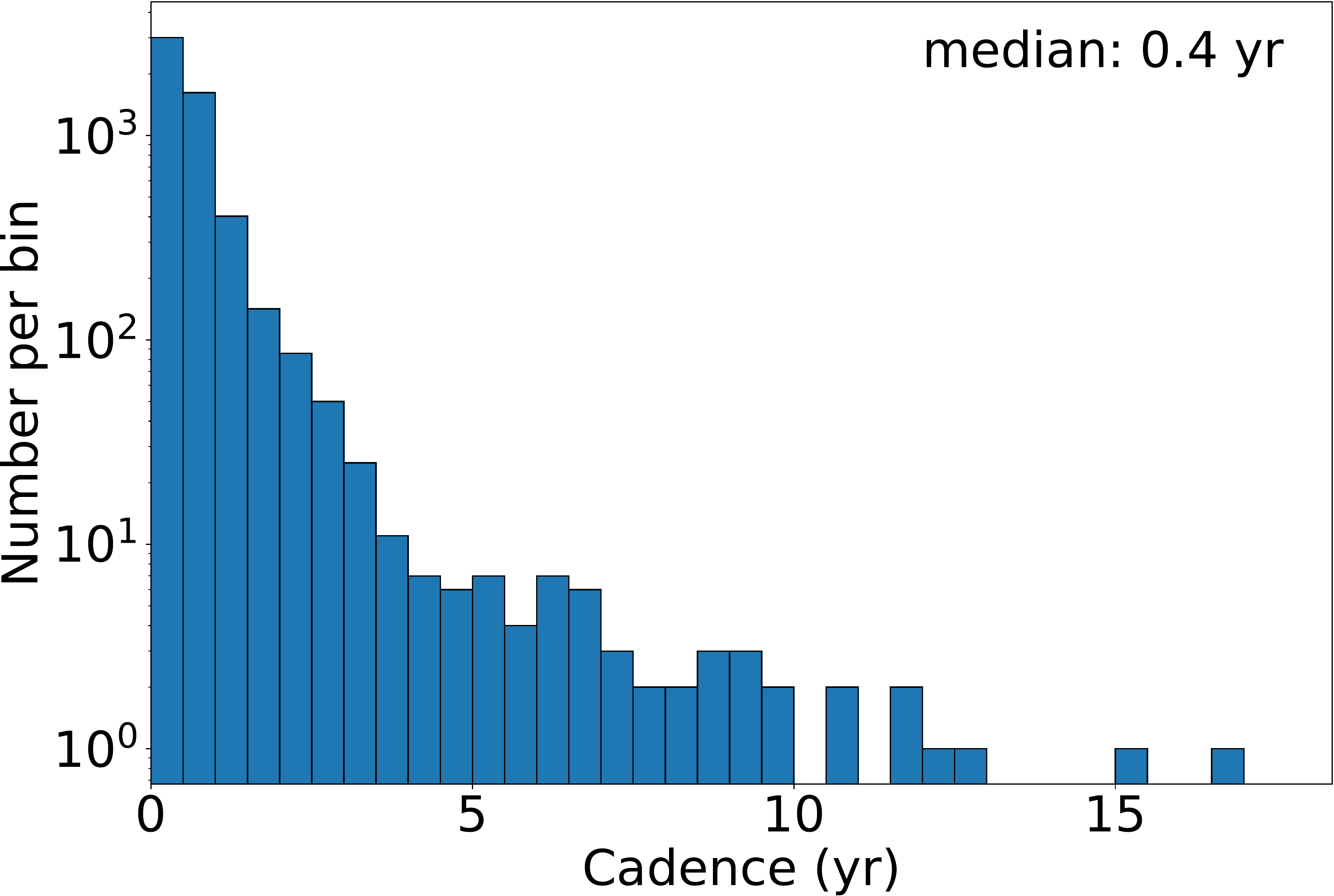}
    \caption{Histogram of the distribution of the time intervals between successive VLBA epochs for all sample sources.
    \label{f:cadence}
    }
\end{figure}

Although the time intervals between successive observations (Fig.~\ref{f:cadence}) varies widely from several months up to 17~years, the peak and the median sampling interval are about half a year. In the MOJAVE programme, the cadence for a source is chosen in such a way that the sources with faster morphological changes are observed more frequently. The quasar 0306+102 (15 epochs) and the BL~Lac 1147+245 (eight epochs) have one epoch at 17 and 15 years apart from the other epochs, respectively. This does not strongly affect the variability estimates as typical variations are measured. The redshift is known for 357 AGNs (259 quasars, 74 BL~Lacs and 24 radio~galaxies) in our sample (\autoref{t:sample_properties}). The redshifts range from 0.05 to 3.40 for the quasars, from 0.02 to 1.67 for BL~Lacs, and from 0.004 to 0.266 for radio galaxies. The median redshift of the sample is 0.79.

\section{Variability and median maps} \label{sec:maps}
We used the MOJAVE programme data and the archival VLBA observations at 15~GHz performed between 1996 January 19 and 2019 August 4. In total, this amounted to 5846 datasets for individual sources at various epochs. The data reduction and imaging process are described in \citet{Lister_2009}. The maps for 419 AGNs have a field of view of 40~mas$\times$40~mas. The remainder were imaged with a larger field to account for their extended structure. All images were gridded with a pixel size of 0.1~mas. The maps of the Stokes parameters $I$, $Q$ and $U$ were restored with a circular Gaussian beam with a FWHM beam size equal to the arithmetic average of the beam major and minor axes, which were estimated using the procedure described in Appendix \ref{appendix:beam_fits}. For better tracing of the brightest parts of the jet, we constructed the total intensity ridgeline, with each point being the intensity-weighted average over the azimuthal slice at a given distance from the core \citep{Pushkarev_2017}. The azimuthal slices were taken with a step of one-pixel size. A circular restoring beam was applied to remove a possible ridgeline dependency on the beam shape if it is highly elongated, thus influencing the ridgeline by the direction of the major axis of the beam. The comparison of the maps convolved with a circular beam with the maps convolved with a mean elliptical beam over the epochs showed that the change of the beam shape does not have a strong influence on the data (the median difference is less than 1~per~cent). The contrast depends on source declination: the median difference is maximum for the AGNs with declination $< 0^\circ$ (up to 4~per~cent for EVPA standard deviation, 6~per~cent for median fractional polarization and 2~per~cent for relative fractional polarization variability), and it decreases with the declination increase. The maps for different epochs were then aligned on the VLBI core position found by the structure modelling in the ($u$, $v$)-plane by means of the \texttt{modelfit} procedure in the Caltech \texttt{Difmap} package \citep{Shepherd_1997}. The modelling approach is described in \citet{Lister_2021}. In our study, the core position was assumed to be stable with time. However, small changes of its position might be produced by the component emergence. The noise levels  $I_{\mathrm{rms}}$, $Q_{\mathrm{rms}}$ and $U_{\mathrm{rms}}$ for the corresponding individual maps were estimated in four outer quadrants, each of 1/25 of the image area. Then the median of the three smallest rms values was taken as an rms over the map. A four-rms-level cutoff was chosen for $I$ images as this is the minimum level at which there are few noise contours in maps.

The next step was to construct individual images of the linearly polarized intensity $P = \sqrt{Q^2 + U^2}$ and the electric vector position angle $\textrm{EVPA} = 0.5\arctan{(U/Q)}$ which lies in a range from $-\pi/2$ to $+\pi/2$. The noise of $P$ of the individual images ($P_{\mathrm{rms}}$)  was estimated as $(Q_{\mathrm{rms}} + U_{\mathrm{rms}})/2$ \citep{Hovatta_2012}. We blanked the $P$ maps at the $3.4P_{\mathrm{rms}}$ level because $P$ follows a Rayleigh distribution, and the $3.4$ rms level corresponds to a $3\sigma$ cutoff for this distribution. Also, pixels with total intensity higher than $4I_{\mathrm{rms}}$ were blanked for further calculations of fractional polarization. The same requirements of blanking were applied to the EVPA images.

Due to a non-Gaussian distribution of polarization noise, the individual $P$ maps are subject to a Rician bias \citep{Wardle}. It increases the observed $P$, especially for the low signal-to-noise ratio regions, while for the high signal-to-noise ratio areas, the difference between the observed and true values can be neglected. We corrected the individual $P$ maps for this bias following \citet{Wardle}
\begin{equation}
    P_{\mathrm{true}} = P_{\mathrm{obs}}\sqrt{1-(P_{\mathrm{obs,std}}/P_{\mathrm{obs}})^2},
\end{equation}
where $P_{\mathrm{true}}$ and $P_{\mathrm{obs}}$ are the corrected and observed polarization intensity, respectively, $P_{\mathrm{obs,std}}$ is the standard deviation evaluated as an average between the $Q$ and $U$ standard deviation, which was estimated following the same procedure as $Q_{\mathrm{rms}}$ and $U_{\mathrm{rms}}$. For a 3$\sigma$ cutoff, the difference between the observed and true values of the polarization intensity is up to 6~per~cent. A more detailed analysis of the Rician bias influence on the polarization data can be found in \cite{MOJAVE_XXI}. It is important to blank the individual $P$ images first and then correct them for the bias because the latter changes the noise distribution, so our estimate of $P_{\mathrm{rms}}$ as an average between $Q_{\mathrm{rms}}$ and $U_{\mathrm{rms}}$ would be incorrect for a Rayleigh distribution. We then constructed the fractional polarization maps $m = P/I$.

For each source in our sample, we constructed images of the median $P$ and $m$ over the epochs in each pixel (median maps $P_{\mathrm{med}}$ and $m_{\mathrm{med}}$) to derive typical values over observations. The moving and quasi-stationary components could influence the median polarization map. In particular, this leads to a non-monotonic distribution of median values along the jet (see Subsection \ref{subsec:var_along}). The EVPA and $m$ variability images were produced using the standard deviation of the values for the individual epochs in each pixel. For the EVPA standard deviation, circular statistics \citep{MARDIA} were used to treat the $\pi$ ambiguity. Circular statistics is a sub-field of statistics which takes into account the periodic behaviour of angular values. The circular standard deviation $\sigma_\mathrm{EVPA}$ was calculated as $\sqrt{-2\ln{R}}$ \citep{CircStat}, where $R$ is the resultant vector length on the unit circle. The circular standard deviation was evaluated using the python scipy statistics library \citep{scipy}, which correctly accounts for the circular range of our EVPA data from $-90^\circ$ to $+90^\circ$. We note that when calculated this way, the circular standard deviation can be arbitrarily large. However, for our data, its values range up to a maximum of about $100^\circ$.

The standard deviation of $m$ and EVPA, $\sigma_m$ and $\sigma_\mathrm{EVPA}$, were assessed only in the pixels unblanked for three or more epochs. As fractional polarization $m$ is positive, its variations have an asymmetric distribution for close to zero $m$ values, and the corresponding $\sigma_m$ likely underestimates the variability. However, $m$ is small mainly in the core area. 
Therefore, $\sigma_m$ can still be used for the analysis. Additionally, we compared the standard deviation with the interquartile range, which is more robust. This range is defined as the difference between the 75th and 25th percentile of the data. It turned out that the interquartile range in the jet is closely matched to the standard deviation and does not show any changes downstream of the outflow. 

\begin{figure*}
    \centering
    \includegraphics[width=\linewidth]{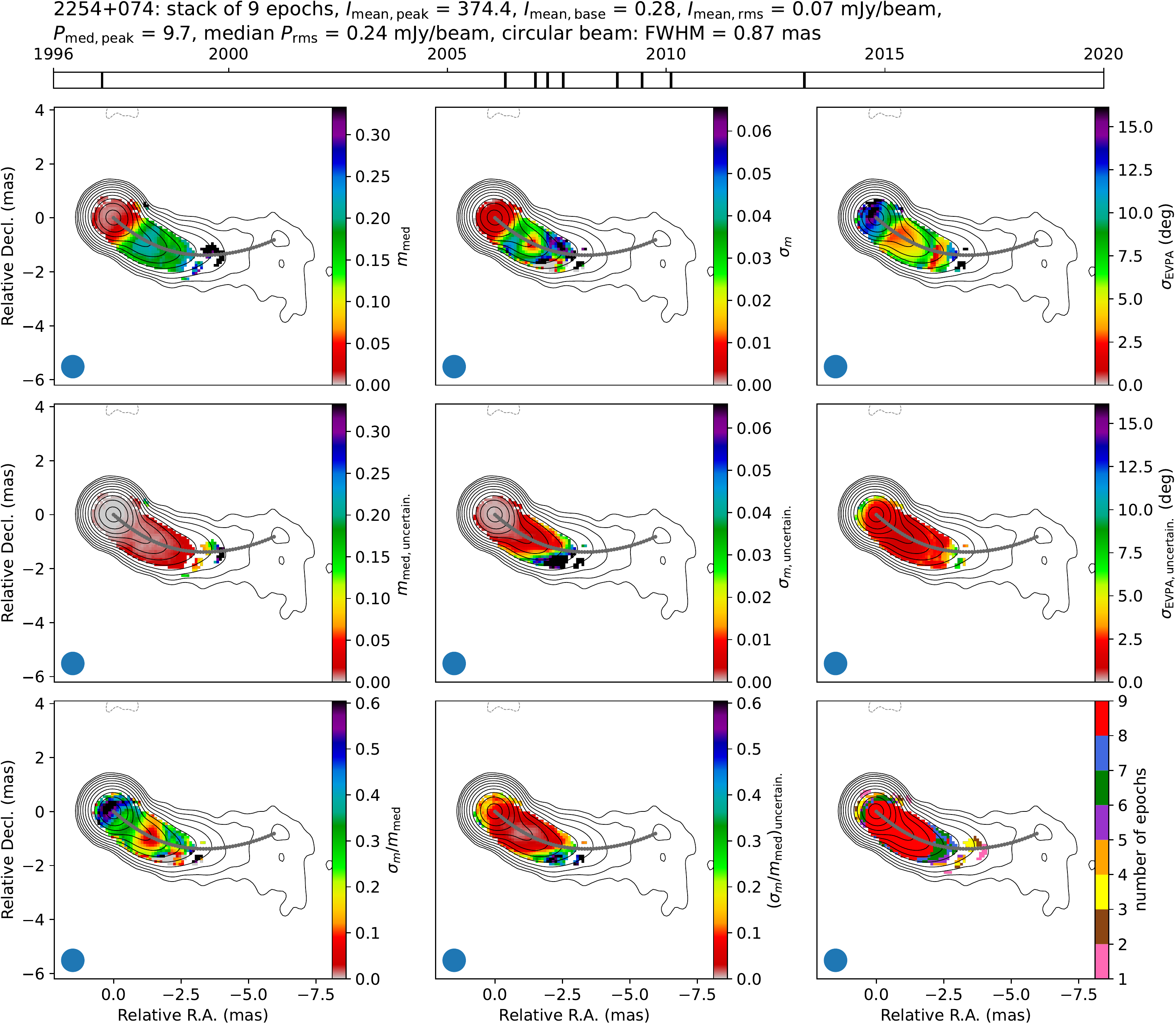}
    \caption{Variability and $m_{\mathrm{med}}$ images. The complete figure set of 436 images is available from the Supplementary data online. Here, we show for guidance the source with nine epochs (the median number of epochs of our sample), 2254+074. The vertical ticks in the top line correspond to the epoch dates. Top row: median fractional polarization $m_{\mathrm{med}}$ (left), fractional polarization standard deviation $\sigma_m$ (centre) and EVPA standard deviation $\sigma_{\mathrm{EVPA}}$ in degrees (right). Middle row: median fractional polarization uncertainty (left), fractional polarization standard deviation uncertainty (centre) and EVPA standard deviation uncertainty (right). Bottom row: relative fractional polarization variability $\sigma_m/m_{\mathrm{med}}$ (left), relative fractional polarization variability uncertainty (centre) and the number of epochs (right) when $m$ and EVPA are unblanked in a given pixel. In all images, the black contours show $I_{\mathrm{mean}}$ contours in increasing powers of two from $4I_{\mathrm{mean, rms}}$ ($I_{\mathrm{mean,base}}$); a single negative contour at level $-I_{\mathrm{mean,base}}$ is shown in gray. The gray dots denote the mean total intensity ridgeline. The distribution of the epochs in the polarization maps is non-uniform. The bottom right image shows the number of epochs when $m$ and EVPA are unblanked in each pixel. In all maps, only the pixels unblanked in three or more Monte Carlo realisations are shown (Appendix \ref{appendix:errors}). The scale range of the colour bars is the same for the value and its uncertainty. The blue circle in the bottom left corner of the images denotes the beam FWHM size.
    \label{f:var_stacked_maps}}
\end{figure*}

\begin{figure*}
    \centering
    \includegraphics[width=\linewidth]{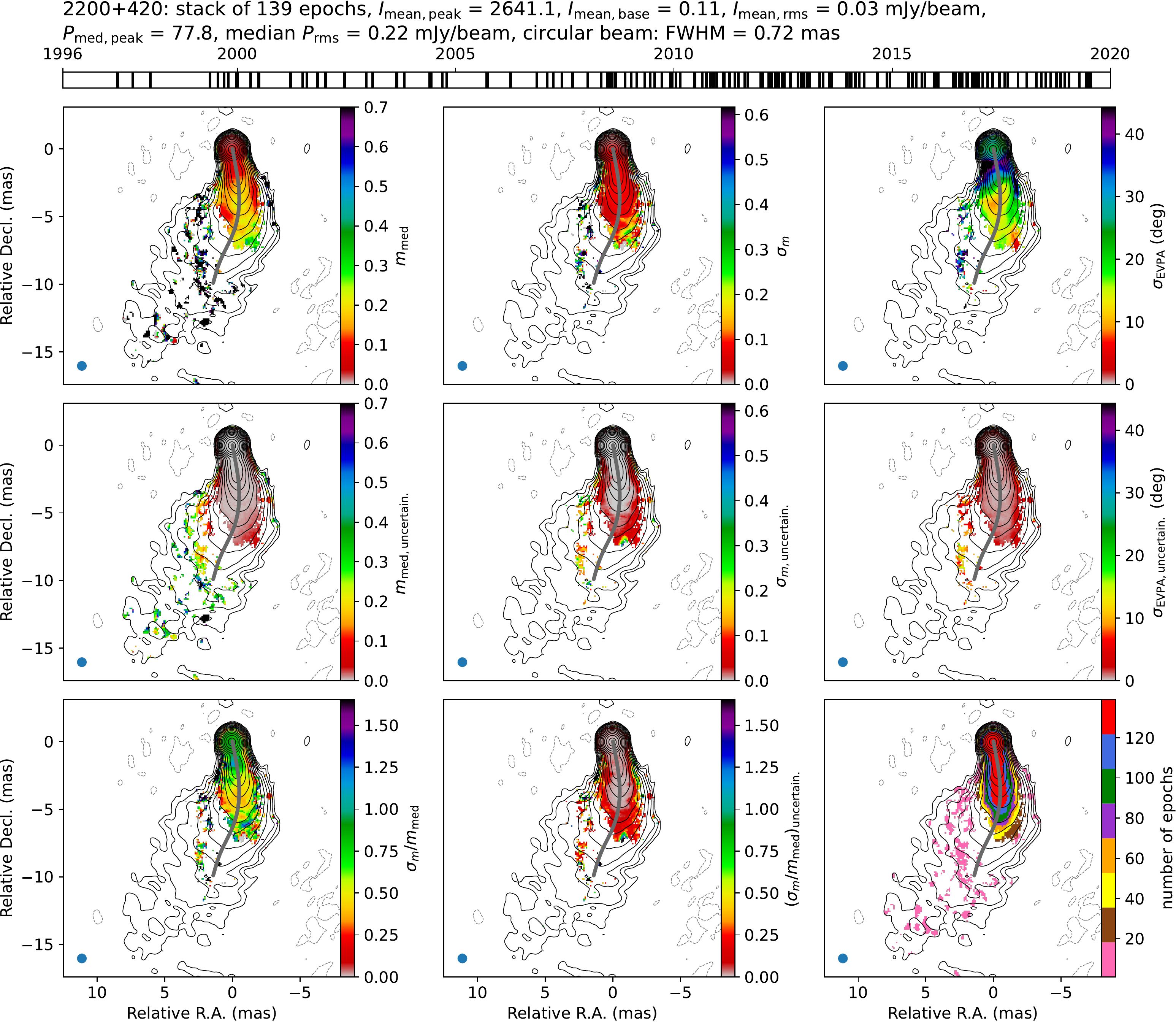}
    \caption{ Variability and $m_{\mathrm{med}}$ images of 2200+420, a source with a maximum number of epochs.  The vertical ticks in the top line correspond to the epoch dates. Top row: median fractional polarization $m_{\mathrm{med}}$ (left), fractional polarization standard deviation $\sigma_m$ (centre) and EVPA standard deviation $\sigma_{\mathrm{EVPA}}$ in degrees (right). Middle row: median fractional polarization uncertainty (left), fractional polarization standard deviation uncertainty (centre) and EVPA standard deviation uncertainty (right). Bottom row: relative fractional polarization variability $\sigma_m/m_{\mathrm{med}}$ (left), relative fractional polarization variability uncertainty (centre) and the number of epochs (right) when $m$ and EVPA are unblanked in a given pixel. The gray dots denote the mean total intensity ridgeline. The blue circle in the bottom left corner of the images denotes the beam FWHM size.
    \label{f:var_stacked_maps_2200+420}}
\end{figure*}

A correction for the bias produced by the CLEAN algorithm (see Appendix \ref{appendix:errors} and \cite{MOJAVE_XXI} for more details) was applied to the $P_\mathrm{med}$, $m_\mathrm{med}$ and $\sigma_m$ maps; $\sigma_{\mathrm{EVPA}}$ was found to be weakly affected by the bias \citep{MOJAVE_XXI}. The CLEAN bias correction leads to negative values for $m_\mathrm{med}$ and $\sigma_m$ in some pixels, numbering less than 3~per~cent of the unblanked pixels for nearly all of the sample sources. Moreover, the vast majority of negative values differ insignificantly from zero. Hence, all negative values of $m_{\mathrm{med}}$ and $\sigma_m$ images were set to zero. 
 
We also constructed mean total intensity $I_{\mathrm{mean}}$ maps (mean $I$ over epochs in each pixel), following \citet{Pushkarev_2017}, and corrected them for the CLEAN bias as well. They found that the stacked maps generally reveal a more complete jet cross-section. 
We note that the $I_{\mathrm{mean}}$ maps play a supporting role in this study, where they are used for ridgeline construction and presentation of the polarization maps. To exclude the noisy pixels from the analysis, the following requirements of blanking were applied: (i) $4I_{\mathrm{mean,rms}}$ on all stacked and standard deviation maps, where $I_{\mathrm{mean,rms}}$ is the noise level of $I_{\mathrm{mean}}$ image and is estimated following the same procedure as $I_{\mathrm{rms}}$, and (ii) $3.4\times\mathrm{median}(P_{\mathrm{rms}})$ on median and standard deviation maps, where $P_{\mathrm{rms}}$ is the rms of the polarization intensity of the individual epochs. The maps of $\sigma_m/m_{\mathrm{med}}$ were produced to characterise relative fractional polarization variability.

After all blankings were made, 11 AGNs had no unblanked pixels, i.e. they are essentially unpolarized down to the level of our sensitivity. Additionally, there are 11 AGNs with only a few polarization pixels unblanked in one or two epochs where poor leakage term solutions created possibly spurious polarization. All of these sources are listed in \autoref{t:unpolarized_agns}. In \cite{MOJAVE_XXI}, multi-epoch mean Stokes parameter maps are used to construct $m_{\mathrm{mean}}$ instead of individual epoch $m$ maps we used to derive $m_{\mathrm{med}}$. The images of $m_{\mathrm{mean}}$ have more unblanked pixels than the $m_{\mathrm{med}}$ maps. The polarization cutoff of $m_{\mathrm{mean}}$ is at $4P_{\mathrm{mean,rms}}$ level, where $P_{\mathrm{mean,rms}}$ is the noise estimate of the mean total intensity map $P_{\mathrm{mean}}$. For $m_{\mathrm{med}}$, the cutoff is at $3.4\times\mathrm{median}(P_{\mathrm{rms}})$ level. The noise of $P_{\mathrm{mean}}$ image is lower than the rms of the polarization intensity of the individual $P$ maps as noise decreases while averaging. Consequently, the cutoff of $m_{\mathrm{med}}$ is higher than that of $m_{\mathrm{mean}}$. If one follows the approach of \cite{MOJAVE_XXI}, the number of sources with no detected polarization is 23. There are two AGNs (0710+196 and 1722+401) in which an EVPA rotation is $\approx 90^{\circ}$ in the core region leading to a cancellation of polarization for the method based on the mean Stokes parameters, while for our approach, $m_{\mathrm{med}}$ is detected in those two sources as there is significant fractional polarization at some epochs. Quasar 0742+103 turned out to be unpolarized in our method because of the bias of $P_{\mathrm{med}}$; after correction for this bias, all unblanked pixels became negative.

\begin{table*}
\caption{Unpolarized AGNs.
\label{t:unpolarized_agns}
}
\centering
\begin{threeparttable}
\begin{tabular}{llccc}
\hline\hline\noalign{\smallskip}
Source & Alias & Opt. class & SED Peak Class & $z$ \\
 (1) & (2) & (3) & (4) & (5) \\
\hline
0055+300 & NGC 315 & G & LSP & 0.017 \\
0111+021 & UGC 00773 & B & LSP & 0.047 \\
0128+554 & TXS 0128+554 & G & \nodata & 0.036 \\
0238$-$084 & NGC 1052 & G & LSP & 0.005 \\
0329+654 & TXS 0329+654 & B & HSP & \nodata \\
0615$-$172 & IVS B0615$-$172 & B & ISP & 0.098 \\
0646+600 & S4 0646+60 & Q & LSP & 0.455 \\
0742+103 & PKS B0742+103 & Q & LSP & 2.624 \\
1128$-$047 & PKS 1128$-$047 & G & LSP & 0.266 \\
1331+170 & OP 151 & Q & LSP & 2.085 \\
1404+286 & OQ 208 & G & LSP & 0.077 \\
1413+135 & PKS B1413+135 & B & LSP & 0.247 \\
1509+054 & PMN J1511+0518 & G & LSP & 0.084 \\
1637+826 & NGC 6251 & G & LSP & 0.024 \\
1833+326 & 3C 382 & G & ISP & 0.058 \\
1845+797 & 3C 390.3 & G & LSP & 0.056 \\
1957+405 & Cygnus A & G & LSP & 0.056 \\
2013$-$092 & PMN J2016$-$0903 & B & ISP & \nodata \\
2021+614 & TXS 2021+614 & G & LSP & 0.227 \\
2031+216 & 4C +21.55 & Q & LSP & 0.174 \\
2043+749 & 4C +74.26 & Q & \nodata & 0.104 \\
2047+098 & PKS 2047+098 & B & LSP & 0.226 \\
\hline
\end{tabular}
\begin{tablenotes}
\item
Columns are as follows:
(1) Source name;
(2) Alias;
(3) Optical Class: quasar (Q), BL~Lac (B), radio~galaxy (G);
(4) Class of Synchrotron Peak Position in SED;
(5) Redshift.
\end{tablenotes}
\end{threeparttable}
\end{table*}

The details of the observations and the properties of the stacked maps corrected for the CLEAN bias are listed in \autoref{t:sample_obs_stack_properties}. Fig.~\ref{f:var_stacked_maps} shows these maps and also variability images for 2254+074 as an example. This AGN has nine epochs, the median number of epochs of our sample. In Fig.~\ref{f:var_stacked_maps_2200+420}, polarization variability and median images are given for 2200+420, the source with the maximum number of epochs. The uncertainties of $m_{\mathrm{med}}$, $\sigma_m$ and $\sigma_\mathrm{EVPA}$ were estimated, as described in Appendix~\ref{appendix:errors}. We estimated $\sigma_m/m_{\mathrm{med}}$ uncertainty as

\begin{small}
\begin{equation}
(\sigma_{m}/m_{\mathrm{med}})_\mathrm{uncertain.} = \sqrt{\bigg(\frac{\sigma_{m,\mathrm{uncertain.}}}{m_{\mathrm{med}}}\bigg)^2 + \bigg(\frac{\sigma_m m_{\mathrm{med, \mathrm{uncertain.}}}}{m_{\mathrm{med}}^2}\bigg)^2}.
\label{eq:std_m_to_m_median_error}
\end{equation}
\end{small}

\begin{table*}
\caption{Observational and multi-epoch properties of the AGN sample.
\label{t:sample_obs_stack_properties}
}
\centering
\begin{threeparttable}
\begin{tabular}{lcccccccc}
\hline\hline\noalign{\smallskip}
Source & First epoch & $\tau$ & $N$ & $b$ & $I_\mathrm{mean,peak}$& $I_\mathrm{mean,rms}$ & $P_\mathrm{med,peak}$ & median $P_{\mathrm{rms}}$\\
&& (yr) && (mas) & (mJy/beam) & (mJy/beam) & (mJy/beam) & (mJy/beam)\\
(1) & (2) & (3) & (4) & (5) & (6) & (7) & (8) & (9)       \\
\hline
0003$-$066 & 2003--02--05 & 9.9 & 18 & 0.93 & 1242.82 & 0.05 & 85.77 & 0.21 \\ 
0003+380 & 2006--03--09 & 7.5 & 10 & 0.73 & 462.12 & 0.06 & 5.64 & 0.20 \\ 
0006+061 & 2011--12--29 & 1.4 & 5 & 0.88 & 149.60 & 0.07 & 6.09 & 0.20 \\ 
0007+106 & 2004--02--11 & 9.4 & 13 & 0.86 & 869.31 & 0.05 & 4.98 & 0.18 \\ 
0010+405 & 2006--04--05 & 5.3 & 12 & 0.72 & 508.14 & 0.06 & 3.05 & 0.17 \\ 
\hline
\end{tabular}
\begin{tablenotes}
\item
Columns are as follows:
(1) Source name;
(2) First epoch observed;
(3) Time coverage $\tau$~(yr);
(4) Number of epochs $N$;
(5) Beam FWHM size $b$;
(6) $I_\mathrm{mean}$~peak~(mJy/beam);
(7) $I_\mathrm{mean}$~rms~(mJy/beam);
(8) $P_\mathrm{med}$~peak~(mJy/beam);
(9) Median epoch $P$~rms~(mJy/beam).
This table is available in its entirety in a machine-readable form in the Supplementary data online and at the CDS VizieR. The first five entries are shown here for guidance.
\end{tablenotes}
\end{threeparttable}
\end{table*}

\section{Results and discussion} \label{sec:results}

\subsection{Core and jet polarization direction variability} \label{subsec:EVPA_var}

\begin{figure*}
    \centering
    \includegraphics[width=\linewidth]{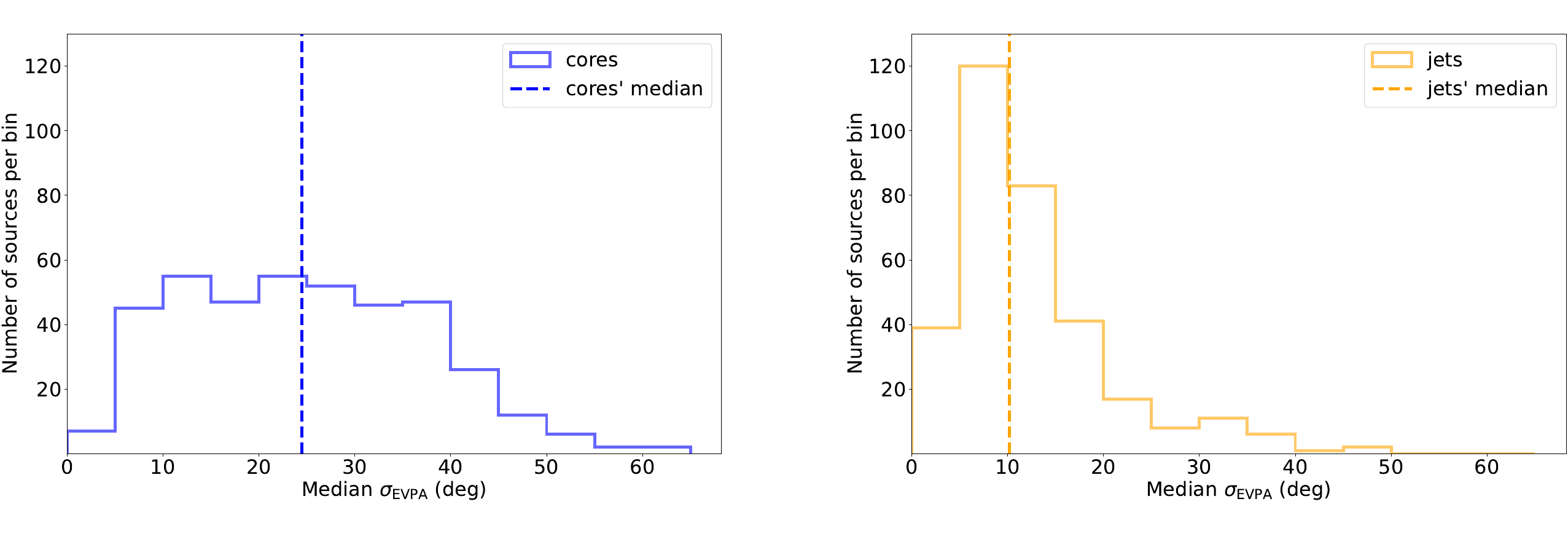}
    \includegraphics[width=\linewidth]{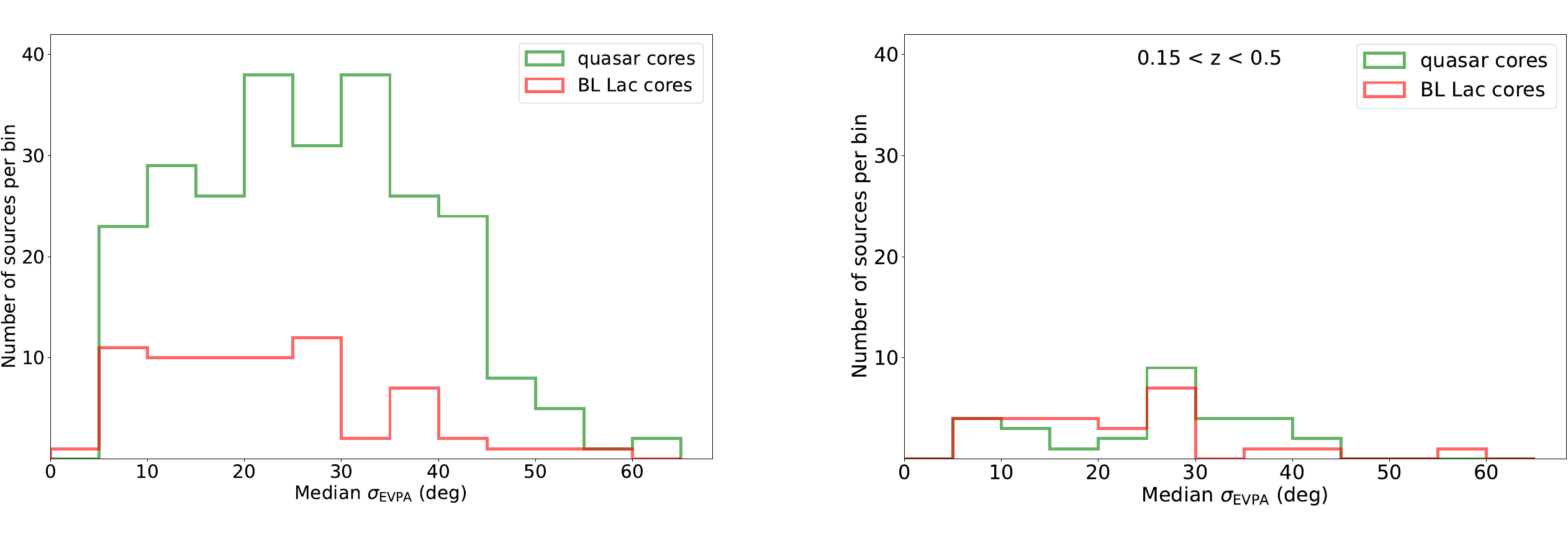}
    \caption{Histograms of median $\sigma_{\mathrm{EVPA}}$ of the cores and of the jets for the whole sample (top row). Histograms of median $\sigma_{\mathrm{EVPA}}$ of quasar and BL~Lac cores (bottom left) and of quasar and BL~Lac cores with a comparable redshift in a range from 0.15 to 0.5 (bottom right). The dashed line in the histograms on the top row shows the median EVPA standard deviation.    
    \label{f:std_EVPA}
    }
\end{figure*}

We investigated the EVPA variability properties of the linearly polarized emission in the core and jet regions. The Gaussian modelfitted core component size averaged over the epochs and convolved with the beam was taken as the FWHM core size on the image. We estimated the distance from the $I_{\mathrm{mean}}$ map peak along the ridgeline at which the core contributes less than 50~per~cent to the observed $P_\mathrm{med}$ to divide the source polarization structure into the core and jet regions. The core was considered as a one-dimensional Gaussian function with the FWHM of the FWHM core size, while the maximum and its position are the peak of $P_\mathrm{med}$ map and its position, respectively. The jet polarized emission dominates that of the core, starting from 1.5 times the FWHM core size on the image from the map centre. We set the boundary between the core and the jet to be perpendicular to the ridgeline at that distance. The median $\sigma_{\mathrm{EVPA}}$ over the core and the jet regions were calculated. Their distributions are shown in Fig.~\ref{f:std_EVPA}. We used the median as a measure of the EVPA variations over the jets because, on average, $\sigma_{\mathrm{EVPA}}$ does not show systematic tendencies along and across the ridgeline (see Subsections \ref{subsec:var_along} and \ref{subsec:var_across}). The core regions have a wide spread of median $\sigma_{\mathrm{EVPA}}$ distributed up to  $\approx 60^\circ$ with the median at $\sim 25^\circ$. For the jets, the distribution is much narrower: the interquartile range is $9^\circ$, compared to $20^\circ$ for the cores. The maximum and the median values for the jets are $\approx 50^\circ$ and about $10^\circ$, respectively. The non-parametric Kendall’s $\tau$ correlation test \citep{Kendall_test} does not show any significant correlations with a redshift for the whole sample or for the optical and synchrotron-peak classes separately. For the analyses in this paper, we assumed the dependence to be significant if the $p$-value is lower than 0.05. Qualitatively, the same results were found for the overlap between the complete flux density limited sample MOJAVE 1.5~Jy Quarter Century \citep[][]{Lister_2019} and our sample. We will refer to the overlap between the two samples as the MOJAVE 1.5JyQC sub-sample. This sub-sample consists of 207 AGNs (about 90~per~cent of the MOJAVE 1.5~Jy Quarter Century sample; about 47~per~cent of our sample). About 76~per~cent of the sources of this sub-sample are LSP quasars, 18~per~cent of the sources are BL~Lacs, the rest of the sources are radio galaxies, optically unidentified sources and a narrow-line Seyfert~1 galaxy. The fraction of LSP BL~Lacs relative to all BL~Lacs in the MOJAVE 1.5JyQC sub-sample is 78~per~cent. It is larger in comparison to our sample (\autoref{t:opt_sed_class}).

 The $k$-sample Anderson-Darling test \citep{AD_test} indicates significant differences between the core and the jet EVPA standard deviation both for the whole sample and separately for the quasars, BL~Lacs, LSP sources (70~per~cent of which are quasars), ISP sources (80~per~cent of which are BL~Lacs). Significant differences were also found for the LSP quasars, LSP BL~Lacs, ISP BL~Lacs and a combination of ISP and HSP (ISP+HSP) BL~Lacs. About 60~per~cent of ISP+HSP BL~Lacs are ISP sources, which drives the significance of the core--jet difference for the combination. 
 
 We verified our results by comparing them with a randomization test \citep{randomization_test}. For a randomization test, we made 10\,000 resamplings from the combination of the $\sigma_{\mathrm{EVPA}}$ sets of the core and the jet. The $p$-value was estimated as the proportion of the resampled data sets which have the Anderson-Darling test statistic larger than that for the original data set. The randomization test $p$-values turned out to be very close to that of the Anderson-Darling test for the initial sets. \autoref{t:std_EVPA_results} presents the results of the randomization test for $\sigma_{\mathrm{EVPA}}$ of our sample for various optical and SED classes. The results for the MOJAVE 1.5JyQC sub-sample are consistent with these results.

\begin{table*}
\caption{Randomization test results on core and jet $\sigma_{\mathrm{EVPA}}$.
\label{t:std_EVPA_results}
}
\centering
\begin{threeparttable}
\begin{tabular}{lcrc}
\hline\hline\noalign{\smallskip}
Test & Result & $P$-value & Number of Sources of Each Type \\
(1) & (2) & (3) & (4) \\
\hline
cores vs jets (whole sample) & Y & <0.0001 & 402 vs 328 \\
quasar cores vs quasar jets & Y & <0.0001 & 251 vs 211 \\
BL~Lac cores vs BL~Lac jets & Y & <0.0001 & 126 vs 98 \\
LSP cores vs LSP jets & Y & <0.0001 & 338 vs 278 \\
ISP cores vs ISP jets & Y & 0.0001 & 39 vs 32 \\
LSP quasar cores vs LSP quasar jets & Y & <0.0001 & 244 vs 207 \\
LSP BL~Lac cores vs LSP BL~Lac jets & Y & 0.0003 & 72 vs 55 \\
ISP BL~Lac cores vs ISP BL~Lac jets & Y & 0.0003 & 33 vs 27 \\
ISP+HSP BL~Lac cores vs ISP+HSP BL~Lac jets & Y & <0.0001 & 54 vs 43 \\
\hline
quasar cores vs BL~Lac cores & Y & <0.0001 & 251 vs 126 \\
\hline
quasar jets vs BL~Lac jets & N & 0.9995 & 211 vs 98 \\
\hline
\end{tabular}
\begin{tablenotes}
\item
Columns are as follows:
(1) Test for the difference of median EVPA standard deviation between two sub-samples;
(2) Result of the test for our sample: Y - difference is significant, N - difference is insignificant;
(3) $P$-value of the result; 
(4) Number of sources in each sub-sample of our sample used in the comparison.
For the MOJAVE 1.5JyQC sub-sample, the results of the tests listed in this Table are consistent with those for the full sample.

\end{tablenotes}
\end{threeparttable}
\end{table*}

The reason for higher $\sigma_{\mathrm{EVPA}}$ values of the core might be blending of several components with appreciably different EVPAs within the unresolved core region. For instance, if the EVPAs of components are aligned with the jet and the jet is bent, then the components have different EVPAs on the sky and the blending produces a change in the polarization direction. The blending of the components decreases fractional polarization as well \citep[e.g.][]{Lister_2005,Helmboldt_2007}.  As these unresolved components move and change their opacity over time, the net blending effect on the EVPA and fractional polarization of the core region will also evolve with time. Direct cancellation between sub-components may produce nulls or flips in the net core polarization, and in rare cases, high opacity of individual sub-components may flip the EVPA of that sub-component, although \citet{Wardle_2018} has shown that a purely opacity-induced EVPA flip is very difficult to observe under conditions normally present in AGN cores.

Another factor which may contribute to less stable EVPA in the core is the strong variability of the observed EVPAs due to temporal changes in the Faraday rotation measure \citep[e.g.][]{Lisakov_2021} in the core region. \citet{Hovatta_2012} found that the AGN cores show significantly higher Faraday rotation measures than the jets. It could also provide the difference between typical $\sigma_{\mathrm{EVPA}}$ seen in the two regions. Large EVPA rotations are seen in a model consisting of a steady `jet' and a variable `burst' \citep{Cohen_2020}.

We also study the behaviour of the core EVPA standard deviation as a function of the optical class. \citet{Hodge_2018} found that the polarization direction of the BL~Lac cores is more stable than that of quasars for a smaller sample with shorter time coverage. The Anderson-Darling test confirms a significant difference between the EVPA standard deviation of the quasar and BL~Lac cores for our sample, as can be visually seen in Fig.~\ref{f:std_EVPA} (bottom left): the EVPA of the BL~Lac cores is more stable than that of quasars. The randomization test gives the $p$-value of less than 0.0001. The higher EVPA variability in the quasar cores can be caused by, on average, higher Faraday rotation measures compared to those in BL~Lacs as found by \citet{Hovatta_2012}. This is also consistent with a statistically lower fractional polarization observed in the quasar cores \citep{MOJAVE_XXI}. The quasars and BL~Lacs selected from a redshift range of $0.1<z<0.5$ do not show any significant difference in $\sigma_{\mathrm{EVPA}}$ of the core regions (Fig.~\ref{f:std_EVPA}, bottom right). This result is driven by two BL~Lacs with the highest median EVPA standard deviation in the core. Consequently, the EVPA standard deviation remains higher for the quasars if sources in a matched redshift range are considered. 

The quasars and BL~Lacs from the LSP-dominated MOJAVE 1.5JyQC sub-sample do not show any difference in the core EVPA standard deviation (the randomization test gives the $p$-value of 0.12). The lack of BL~Lacs with a more stable core EVPA in the MOJAVE 1.5JyQC sub-sample leads to an increase in the median EVPA standard deviation of the BL~Lac core region and makes it comparable to that of a quasar.

\subsection{Variability in EVPA, fractional polarization and $m_{\mathrm{med}}$ along the ridgeline} \label{subsec:var_along}

We analysed the statistics of jet variability along the outflow ridgeline. The sources were divided into the core and the jet regions, following the same procedure as in Subsection~\ref{subsec:EVPA_var}. We studied the changes of $m_{\mathrm{med}}$, $\sigma_m/m_{\mathrm{med}}$ and $\sigma_{\mathrm{EVPA}}$ along the jet ridgelines beyond the core region. The fractional polarization standard deviation $\sigma_m$ was not considered because it was found to significantly correlate with the median fractional polarization. This dependence might occur due to the influence of noise at the low signal-to-noise regions. In the inner jet, the correlation between $m_{\mathrm{med}}$ and $\sigma_m$ might be intrinsic. It could be explained by means of the magnetic field produced by turbulent cells \citep{Marscher_2014}: higher observed fractional polarization could be due to a smaller number of independent magnetic field regions within the beam or along the line of sight. The low values of $\sigma_m/m_{\mathrm{med}}$ would then correspond to the locations in the jet with less influence from stochastically changing fields. Relative variability $\sigma_m/m_{\mathrm{med}}$ is a better measure of magnetic field order changes than $\sigma_m$ because it measures variations relative to the local degree of order as characterised by $m_{\mathrm{med}}$ and is not expected a-priori to depend on the distance from the core, as $\sigma_m$ does. For the correlation analysis, we used the Kendall\textquotesingle s $\tau$ test, considering the $p$-values less than 0.05 to be significant.  It is crucial to consider only independent pixels to avoid false significant dependencies. Thus, a single pixel every half of the beam size along the ridgeline was taken. 

However, the Kendall\textquotesingle s $\tau$ test does not consider the associated uncertainty. Hence, to exclude cases when a correlation is formally significant but the uncertainties are so large that this correlation might be spurious, we eliminated the jets if all analysed pixels along the ridgeline have the same value of the considered quantity within a $1\sigma$ uncertainty. We considered only the jets having five or more polarization-sensitive pixels in their profiles along the ridgeline after all restrictions described above were applied.

\begin{figure*}
    \centering
    \includegraphics[width=\linewidth]{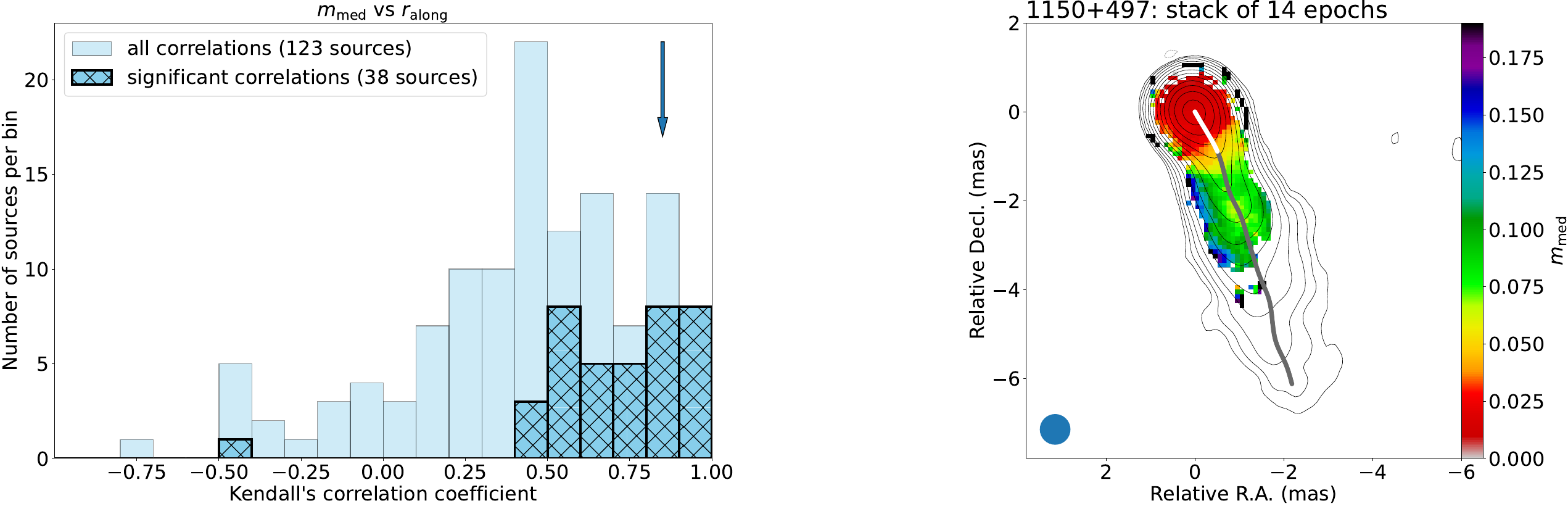}
    \caption{Histogram of the distribution of the Kendall\textquotesingle s correlation coefficients for the median fractional polarization $m_\mathrm{med}$ against the distance along the ridgeline from the core $r_{\mathrm{along}}$ (left) and the $m_\mathrm{med}$ image of 1150+497 (right) as an example of a source showing a positive significant correlation between $m_\mathrm{med}$ (distribution according to the colour bar) and $r_{\mathrm{along}}$.
    The filled bins on the histogram correspond to all considered correlations, whereas the hatched ones correspond to significant correlations, i.e. the coefficients from the hatched histogram are included in the light blue histogram. The arrow shows the bin where the correlation coefficient of 1150+497 lies. On the map, the white and gray lines denote the mean total intensity ridgeline in the core and jet regions, respectively. The blue circle in the bottom left corner of the map denotes the beam FWHM size.
    \label{f:r_along_m_median}
    }
\end{figure*}

\begin{figure}
    \centering
    \includegraphics[width=\linewidth]{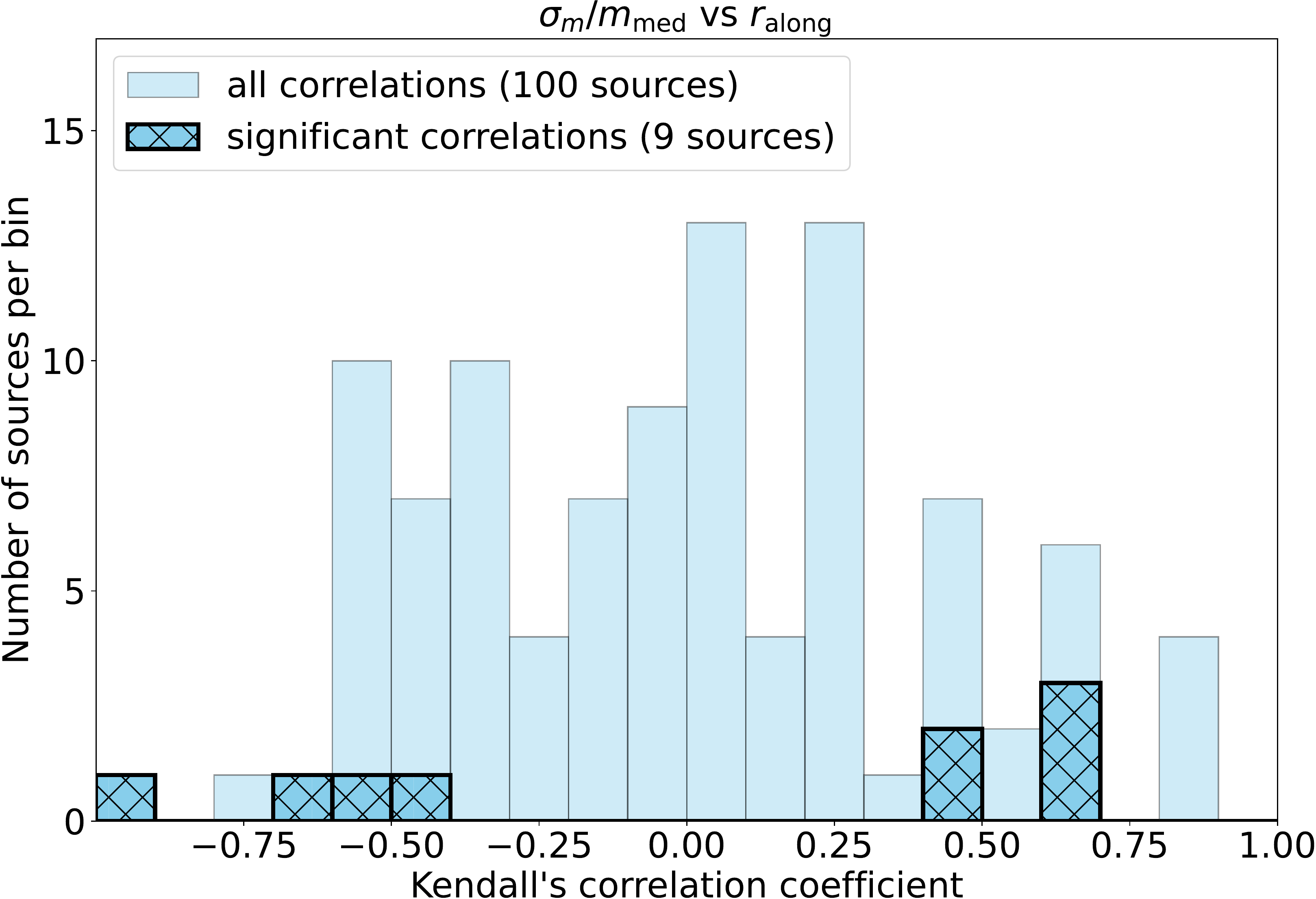}
    \caption{Histogram of the distribution of the Kendall\textquotesingle s correlation coefficients for relative fractional polarization variability $\sigma_m/m_{\mathrm{med}}$ against the distance along the ridgeline from the core $r_{\mathrm{along}}$. The filled bins correspond to all considered correlations, whereas the hatched ones correspond to significant correlations.
    \label{f:r_along_std_m_to_m_median}
    }
\end{figure}

\begin{figure*}
    \centering
    \includegraphics[width=\linewidth]{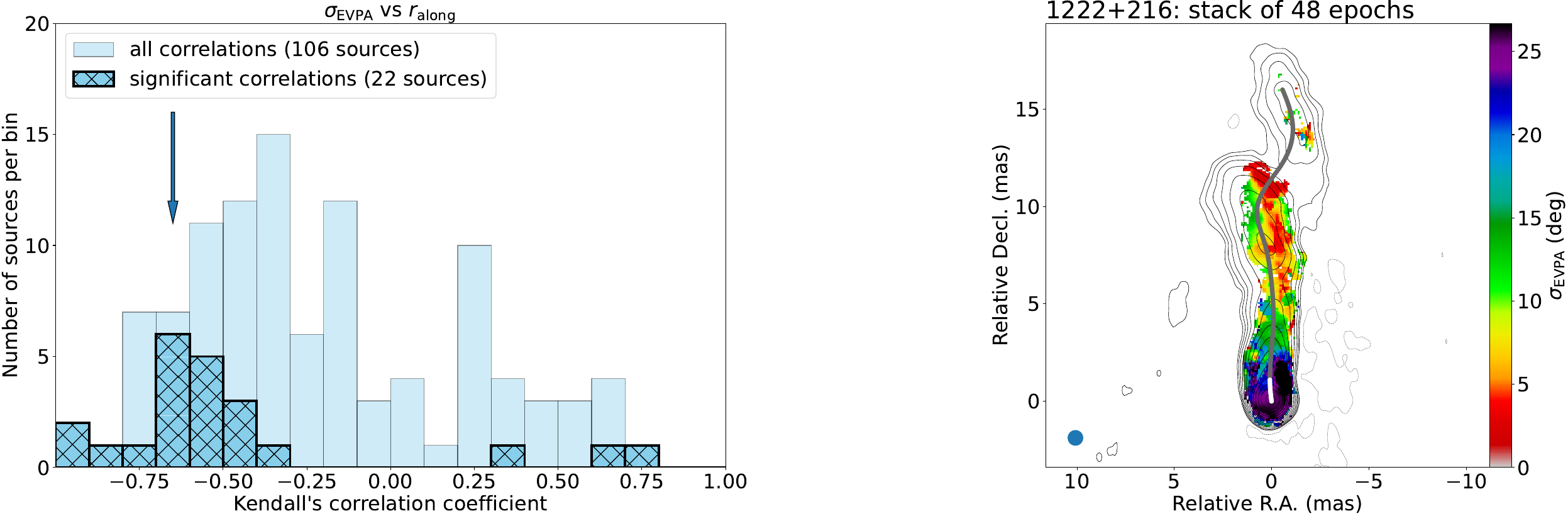}
    \caption{Histogram of the distribution of the Kendall\textquotesingle s correlation coefficients for EVPA standard deviation $\sigma_{\mathrm{EVPA}}$ against the distance along the ridgeline from the core $r_{\mathrm{along}}$ (left) and the $\sigma_{\mathrm{EVPA}}$ image of 1222+216 (right) as an example of a source showing a significant anti-correlation between $\sigma_{\mathrm{EVPA}}$ (distribution according to the colour bar) and $r_{\mathrm{along}}$.
    The filled bins on the histogram correspond to all considered correlations, whereas the hatched ones correspond to significant correlations, i.e. the coefficients from the hatched histogram are included in the light blue histogram. The arrow shows the bin where the correlation coefficient of 1222+216 lies. On the map, the white and gray lines denote the mean total intensity ridgeline in the core and jet regions, respectively. The blue circle in the bottom left corner of the map denotes the beam FWHM size.
    \label{f:r_along_std_EVPA}
    }
\end{figure*}

In Fig.~\ref{f:r_along_m_median}, we present the Kendall\textquotesingle s $\tau$ distribution for $m_\mathrm{med}$ against the distance along the ridgeline from the core $r_\mathrm{along}$. In total, there are 123 AGNs. For 38 of them, there is a significant trend in $m_\mathrm{med}$ with $r_\mathrm{along}$. Almost all the significant trends (correlations), with $p$-values $<0.05$, are positive, i.e. $m_{\mathrm{med}}$ increases with distance from the core. This is also typically seen in the $m$-maps at individual epochs (e.g. \citealt{Cawthorne_1993} at 5~GHz, \citealt{Lister_2005} at 15~GHz, \citealt{Lister_2000} at 22~GHz). In the case of $\sigma_m/m_\mathrm{med}$ (Fig.~\ref{f:r_along_std_m_to_m_median}), only nine jets out of 100 exhibit a significant dependence on $r_{\mathrm{along}}$: five sources show an increase as a function of distance, whereas four AGNs show a decrease. We also analysed a correlation between $\sigma_{\mathrm{EVPA}}$ and $r_{\mathrm{along}}$. A distribution of the corresponding Kendall\textquotesingle s $\tau$ values (Fig.~\ref{f:r_along_std_EVPA}, left) shows that 22 sources out of 106 have a significant, predominantly negative, correlation. In Fig.~\ref{f:r_along_std_EVPA} (right), we present an example of a source with a typical significant tendency. Although the majority of $m_\mathrm{med}$ and EVPA standard deviation trends with $r_\mathrm{along}$ are insignificant, the correlations with $p$-value $<0.05$ have a preferable sign of the coefficient: negative values for $\sigma_{\mathrm{EVPA}}$ and positive values for $m_\mathrm{med}$. Hence, the increase in the EVPA stability and the median polarization degree appear to be typical of AGN jets. Otherwise, the number of significant correlations and anti-correlations would be comparable. AGNs with more extended projected jets tend to have a larger fraction of significant correlations for all considered quantities than AGNs with short projected jets.

We estimated the dependence of the fraction of significant correlations between $\sigma_{\mathrm{EVPA}}$ and $r_{\mathrm{along}}$ on the AGN time coverage for the sources with a comparable epoch number in a range of 10--20. For the time coverage less than 12~years, the fraction of significant correlations is about 0.10; whereas for the full sample, the fraction is 0.21. The sources observed on longer time-scales show the fraction increase with increased time coverage. Hence, the time coverage of at least 12~years is needed to show the EVPA standard deviation trend along the jet. The analysis of the dependence of fraction of significant correlations on the number of epochs for the sources with a time coverage of 10--15~years shows that at least 15--20~epochs is needed to reveal the EVPA standard deviation tendency. The increase in the fraction of significant correlations with the time coverage and the number of epochs gives evidence that the EVPA standard deviation distribution is sensitive to these parameters. A similar analysis for $m_\mathrm{med}$ against $r_{\mathrm{along}}$ shows that there is no dependence of the fraction of significant correlations on either time coverage or epoch number. Therefore, the median fractional polarization remains more or less the same with the increase in the time coverage and the number of observations. About 77~per~cent (17 out of 22) of significant correlations between $\sigma_{\mathrm{EVPA}}$ and $r_{\mathrm{along}}$ are for AGNs with a time coverage of at least 12~years and more than 15 epochs. Consequently, EVPA standard deviation needs enough time and observation epochs to exhibit the trend.

Both bright features and jet bends contribute to the lack of significant correlations in some sources. Appendix \ref{appendix:AGNs} gives notes on several sources which show a significant anticorrelation  of $m_{\mathrm{med}}$, a significant correlation of $\sigma_{\mathrm{EVPA}}$ or a significant dependence of $\sigma_m/m_\mathrm{med}$ with $r_{\mathrm{along}}$.

Near the jet base, low $m_\mathrm{med}$ may be the consequence of strong Faraday depolarization, and high EVPA standard deviation could be the manifestation of component blending and jet bends on small scales. The observed $m_\mathrm{med}$ and $\sigma_{\mathrm{EVPA}}$ maps convolved with the beam reflect the magnetic field distribution and evolution with beam resolution.  Low $m_\mathrm{med}$ and high EVPA standard deviation near the core could reflect a less ordered magnetic field with a less stable direction, whereas the trend of an $m_\mathrm{med}$ increase and $\sigma_{\mathrm{EVPA}}$ drop suggest the growth of order and stability of a magnetic field along the jet. Also, near the jet base, new components with various EVPAs emerge producing low $m_\mathrm{med}$ and high $\sigma_{\mathrm{EVPA}}$ due to not being resolved yet. The jet expands downstream; hence, the components become more extended. This leads to a progressive increase in $m_\mathrm{med}$ and a decrease in the EVPA standard deviation downstream in the jet. Another explanation for the fractional polarization increase could be the change of the helical field pitch angle, as shown by \citet{Porth_2011} in their relativistic MHD simulations of jets.

Both internal and external Faraday rotation can cause depolarization \citep{Burn_1966}. \citet{Hovatta_2012} rewrote the expressions for both cases as $\ln m = \ln m_0 - b\lambda^4$ for the rotation measures not exceeding 800~rad~m$^{-2}$, where $m_0$ is the maximum $m$ in the specific magnetic field configuration, $b$ is $2\mathrm{RM}^2$ for the internal depolarization and $2\sigma^2$ for the external depolarization, where $\sigma$ is the standard deviation of the RM fluctuations, and RM is the rotation measure. Thus, a completely uniform external Faraday rotation would not produce depolarization. The median fractional polarization $m_{\mathrm{med}}$ near the jet base is about 0.04 for our sample, and it increases over the ridgeline up to approximately 0.18, which would require a change in the rotation measure along a ridgeline of about 2200~rad~m$^{-2}$. However, the observed change is about a few hundred~rad~m$^{-2}$ \citep{Hovatta_2012}. A RM value of 400~rad~m$^{-2}$ gives a rise of $m$ of less than 2~per~cent. Consequently, this mechanism can contribute only a little to the observed increase of $m_\mathrm{med}$. The variations of the rotation measure RM can lead to changes of EVPA as $\sigma_{\mathrm{EVPA}} \sim \sigma_{\mathrm{RM}}\lambda^2$. For a typical decrease of $\sigma_{\mathrm{EVPA}}$ from $28^\circ$ in the inner region to $8^\circ$ in the outer region of the jet beyond the core, RM variability drops from 1225~rad~m$^{-2}$ to 350~rad~m$^{-2}$. Other observations give RM variations along the jet of several hundreds~of~rad~m$^{-2}$ \citep[e.g.][]{Zavala_2001, Hovatta_2012, Zamaninasab_2013}.

The distributions of the correlation coefficient  of polarization variability against $r_\mathrm{along}$, which are seen for the whole sample, remain qualitatively the same for quasars, BL~Lac objects, radio~galaxies and LSP, ISP and HSP sources if considered separately. The results for the MOJAVE 1.5JyQC sub-sample are consistent with those for the full sample.

\subsection{Variability in EVPA, fractional polarization and $m_{\mathrm{med}}$ transverse to the ridgeline} \label{subsec:var_across}

We also analysed the behaviour of $m_{\mathrm{med}}$, $\sigma_m/m_{\mathrm{med}}$ and $\sigma_{\mathrm{EVPA}}$ along the slices transverse to the ridgeline for 312 jets whose width exceeded three beam sizes at least in one slice. Following \citet{Hovatta_2012}, this requirement was chosen as a criterion for very well transversely resolved AGN jets. The same  profile selection procedure was applied as in the case of the profiles along the ridgeline (Section \ref{subsec:var_along}). To have a sufficient number of points, both sides of the slice were analysed together against the absolute distance from the ridgeline in the transverse direction. The slices were constructed at each ridgeline pixel, and the distance between the adjacent pixels is 0.1~mas.

\begin{figure}
    \centering
    \includegraphics[width=\linewidth]{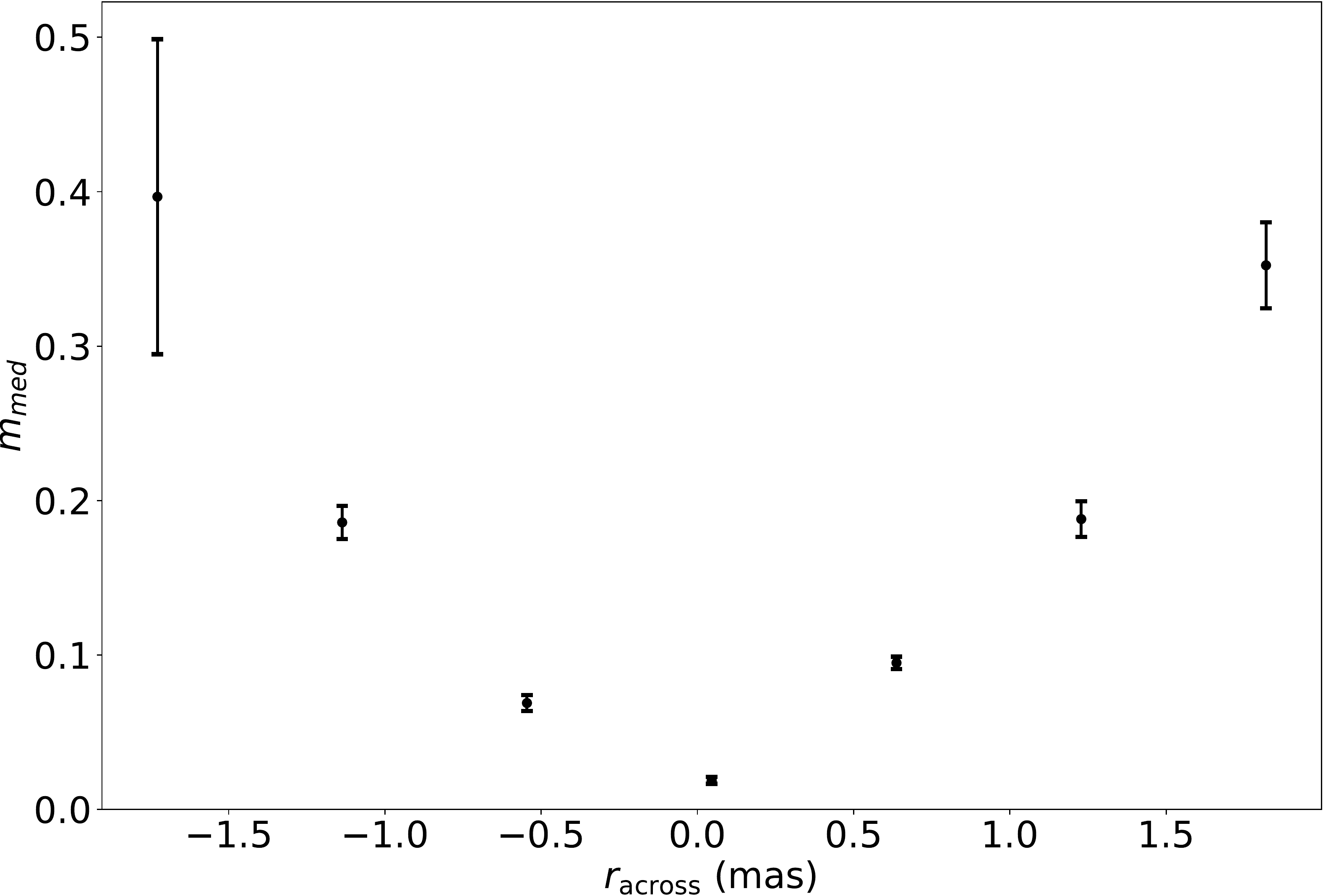}
    \caption{Distribution of $m_\mathrm{med}$ across the jet of 0336$-$019 at 4~mas from the core along the ridgeline. $r_{\mathrm{across}}$ is a distance from the ridgeline to the edges. The Kendall\textquotesingle s $\tau$ test gives a correlation coefficient of 0.9 and a $p$-value of 0.003, i.e. the median fractional polarization significantly increases toward the jet edges. For the test, both sides of the slice are folded over.
    \label{f:r_across}
    }
\end{figure}

For $m_\mathrm{med}$, 82 AGN jets have at least one slice appropriate for this analysis, while 70 of those have two or more slices. The median number of slices per jet is ten. These sources are mostly quasars. About 40 sources have slices with significant trends in $m_\mathrm{med}$ from the ridgeline toward the jet edge ($p$-values for the correlations $<0.05$). We used the Kendall\textquotesingle s $\tau$ test to estimate the significance of the trends. An example of a slice with a significant tendency is given in Fig.~\ref{f:r_across}.  These correlations are predominantly positive, i.e. closer to the jet axis, $m_\mathrm{med}$ is lower than at the jet edge. Our analysis shows that $m_\mathrm{med}$ displays increases toward the jet edges in more than 80~per~cent of cases. Such a distribution could be evidence for a helical magnetic field \citep[e.g.][]{Clausen-Brown_2011}. We detected regions where $m_\mathrm{med}$ is significantly higher at the jet edges than in the jet axis in seven AGNs that show significant transverse rotation measure gradients \citep{Gomez_2011,Hovatta_2012, Gabuzda_2014, Gabuzda_2015, Gabuzda_2017, Gabuzda_2018}: 0430+052, 0735+178, 1226+023, 1611+343, 2037+511, 2200+420, 2230+114. Both the increase in the fractional polarization toward the jet edges and the transverse rotation measure gradient provide evidence of a toroidal or helical magnetic field \citep[e.g. a review paper of ][]{Gabuzda_2021}. A more detailed consideration of this effect together with the corresponding interpretations for more AGNs is given in \cite{MOJAVE_XXI}.

For $\sigma_m/m_{\mathrm{med}}$, only 0454$-$234 (one out of 65 sources) has more than 50~per~cent of significant correlations (six out of ten slices show significant anti-correlation). Hence, the median fractional polarization becomes relatively more stable farther from the ridgeline. In the case of $\sigma_{\mathrm{EVPA}}$, there are two out of 65 AGNs with more than half of significant correlations: 0208+106 and 0738+313. The BL~Lac object 0208+106 has one slice in total, which is significant, and 0738+313 has two slices in total, one of which is significant. The large number of insignificant correlations is explained by the different behaviour of polarization variability and the median fractional polarization at the opposite sides of slices due to, for example, asymmetric interaction with the ambient medium and non-monotonic distribution across the jet. Also, lower signal-to-noise ratio at the jet edges, compared to the ridgeline, can increase the number of insignificant correlations. Consistent results were found for sources comprising the MOJAVE 1.5JyQC sub-sample.

\section{Summary} \label{sec:sum}

We produced and analysed the VLBA images of linear polarization variability ($\sigma_m/m_{\mathrm{med}}$ and $\sigma_{\mathrm{EVPA}}$) and the median fractional polarization $m_{\mathrm{med}}$ of 436 AGNs 
at 15~GHz. The median and standard deviation were chosen as measures of typical values and variability, respectively. The maps were corrected for the Rician and CLEAN biases. They lead to increased fractional polarization in the low signal-to-noise regions. The uncertainties of $\sigma_m$, $\sigma_m/m_{\mathrm{med}}$, $\sigma_{\mathrm{EVPA}}$ and the $m_{\mathrm{med}}$ maps were estimated using Monte Carlo simulations.

The main results of our study are as follows.
\begin{enumerate} 
\item EVPA standard deviation of the core is found to be significantly higher than that of the jet for the whole sample and separately for the sources of different optical and SED classes. The main reason for this could be several components blending and jet bending within the core.

\item For BL~Lac cores, the EVPA is more stable than that of quasars, confirming the earlier finding of \citet{Hodge_2018}. This result remains if we compare BL~Lacs and quasars in a matched redshift range of $0.1<z<0.5$. Lower Faraday rotation measure in BL~Lacs in comparison to quasars \citep{Hovatta_2012} can lead to more stable polarization direction in their cores, suggesting a stronger ordered magnetic field component.

\item We found no correlation with the redshift for the median EVPA standard deviation of the core and the jet for the whole sample and for the quasars, BL~Lacs, LSP, ISP and HSP AGNs if considered separately.

\item For 19 out of 106 AGNs, $\sigma_{\mathrm{EVPA}}$ decreases along the ridgeline down the jet. The median fractional polarization $m_{\mathrm{med}}$ shows a significant, preferential increase with distance from the core along the ridgeline for 38 out of 123 sources. These trends of EVPA standard deviation and $m_{\mathrm{med}}$ are typical of sources as significant correlations have a preferable sign of the correlation coefficient. Otherwise, there would be comparable number of significant correlations and anti-correlations for each of these quantities. The EVPA standard deviation correlations with the jet distance are stronger for observations spanning time-scales longer than 12~years with more than 15~epochs of observations. The reason for high $\sigma_{\mathrm{EVPA}}$ and low $m_{\mathrm{med}}$ in the core could be strong Faraday depolarization, blending of components and jet bending on small scales. On the other hand, a magnetic field which becomes more ordered and stable in direction with increasing separation from the core could explain the EVPA standard deviation decrease and $m_{\mathrm{med}}$ increase down the jet. The change of the helical magnetic field pitch angle along the jet also might give the trend of the $m_{\mathrm{med}}$, as shown in \cite{Porth_2011}. These results remain qualitatively the same if we consider optical and synchrotron peak classes separately.

\item A large number of AGNs have insignificant tendencies of EVPA standard deviation, relative fractional polarization variability and median fractional polarization downstream in the jet. Bright jet features and jet bends might contribute to the complex behaviour.

\item For AGNs with a resolved transverse structure, the vast majority of slices show an increase in the median fractional polarization from the ridgeline toward the edges. This could provide evidence for  helical magnetic field in the jet \citep[e.g.][]{Clausen-Brown_2011, Porth_2011}. The physical scenarios of fractional polarization rising to the jet edges are considered in \cite{MOJAVE_XXI}.

\end{enumerate}

\hfill

\section*{Acknowledgements}

We thank the anonymous referee for helpful suggestions on the manuscript, the members of the MOJAVE team for discussions of the paper, Eduardo Ros for valuable comments on the manuscript as well as Elena Bazanova for language editing.
The analysis presented in this paper was supported by the Russian Foundation for Basic Research grant 20-32-90108.
The estimation of the uncertainties presented in the Appendix~\ref{appendix:errors} was supported by the Russian Science Foundation grant 21-12-00241.
The MOJAVE project was supported by NASA-Fermi grants NNX08AV67G, NNX12A087G and NNX15AU76G. 
This work is part of the M2FINDERS project which has received funding from the European Research Council (ERC) under the European Union’s Horizon 2020 Research and Innovation Programme (grant agreement No 101018682).
This research made use of the data from the University of Michigan Radio Astronomy Observatory (UMRAO), which was supported by the University of Michigan and by a series of grants from the National Science Foundation, most recently No.~AST-0607523 and by a series of grants from the NASA Fermi G.I. Program. This research made use of the NASA/IPAC Extragalactic Database (NED), which is operated by the Jet Propulsion Laboratory, California Institute of Technology, under contract with the National Aeronautics and Space Administration. The Long Baseline Observatory and the National Radio Astronomy Observatory are facilities of the National Science Foundation operated under a cooperative agreement by Associated Universities, Inc. This work made use of the Swinburne University of Technology software correlator \citep{2011PASP..123..275D} developed as part of the Australian Major National Research Facilities Programme and operated under licence.

\section*{Data Availability}

The fully calibrated images of the Stokes parameters $I$, $Q$ and $U$ at 15~GHz from the MOJAVE programme and EVPA standard deviation images are available online\footnote{\url{https://www.cv.nrao.edu/MOJAVE}}. Results of model fitting are taken from \citet{Lister_2021}.

\bibliographystyle{mnras}
\bibliography{references}

\appendix
\section{Estimation of uncertainties} \label{appendix:errors}

To estimate the uncertainties of the stacked as well as individual epoch images of total intensity, linearly polarized flux, EVPA and fractional polarization, as well as their variability measured by the standard deviation of the individual epoch values, we employed Monte Carlo (MC) simulations. \citet{2019MNRAS.482.1955P} found this approach statistically more optimal than the conventional approach of \citet{Hovatta_2012} for individual epoch images. The main idea is to replicate the observed data sets a large number of times and to assess the impact of different noise realisations across data replications on the resulting stacked or individual epoch map.

To obtain the artificial visibility data set at a specific epoch in a single MC realisation, we do the following steps:

\begin{enumerate}
    \item Create model visibilities using $I$, $Q$ and $U$ CLEAN models and $uv$-coverage obtained from the real data set. Here, we shifted the models to put the phase centre to the core position estimated by fitting a Gaussians model to visibility \citep{Lister_2021}.
    \item Add thermal noise estimated from the observed data to the model visibilities. We employed the successive differences approach to estimate the noise at each baseline from the observed data \citep{briggs}.
    \item Model the residual uncertainty of the amplitude scale of the individual antenna gains, scale amplitudes of the parallel and cross-hand correlation on a random factor. For this procedure, draw two random scale factors $C_R$ and $C_L$ (corresponding to $R$ and $L$ antenna polarization) from the Normal distribution $N(0, 0.035)$ and multiplied hand $XY$ on a factor $C_X C_Y$. This corresponds to the 5~per~cent uncertainty estimated by \citet{Hovatta_2014}.
    \item Model the residual polarization leakage (D-terms) uncertainty, draw the real and imaginary part of the residual D-term for each polarization ($R$, $L$) for each IF for $i$-th antenna from a Normal distribution $N(0, \sigma_{\rm D})$, where $\sigma_{\rm D}$ was estimated from the scatter of the D-term solutions of the MOJAVE experiments for each antenna. For non-VLBA antennas (VLA single station at eight epochs), we used the median value of $\sigma_{\rm D}$ for the VLBA antennas. Add the obtained D-terms into the data using linear approximation \citep{Roberts_1994}.
    \item Model the uncertainty of the absolute EVPA of the linear polarization, rotate EVPA to a random value drawn from $N(0, \sigma_{\rm EVPA})$, where $\sigma_{\rm EVPA} = 3^\circ$ was estimated in \citep{Hovatta_2012}.
\end{enumerate}
The generated artificial data sets were imaged in \texttt{Difmap}, using the same script as the observed data. Then the median estimation (for polarized flux and fractional polarization) and the construction of standard deviation maps from the obtained individual epoch images were performed.
We estimated the dispersion (i.e. the random uncertainty) at each single pixel by the scatter of the pixel values across MC realisations. We also estimated the bias (i.e. systematic uncertainty) due to the CLEAN algorithm as a difference between the mean of the images obtained from the MC realisations and the `true' value.  To create `Ground Truth' images, we employed the CLEAN models which were used to create MC realisations (see item (i) in the list above) convolved with a CLEAN beam. The bias estimation is not feasible within the conventional approach \citep{Hovatta_2012}, as it requires knowledge of the true model. A more detailed description of the CLEAN bias and a comparison of the uncertainty estimates with the analytical ones \citep{Hovatta_2012} can be found in a companion paper \citep{MOJAVE_XXI}.

\section{Beam size fits} \label{appendix:beam_fits}

\begin{figure}
    \centering
    \includegraphics[width=\columnwidth]{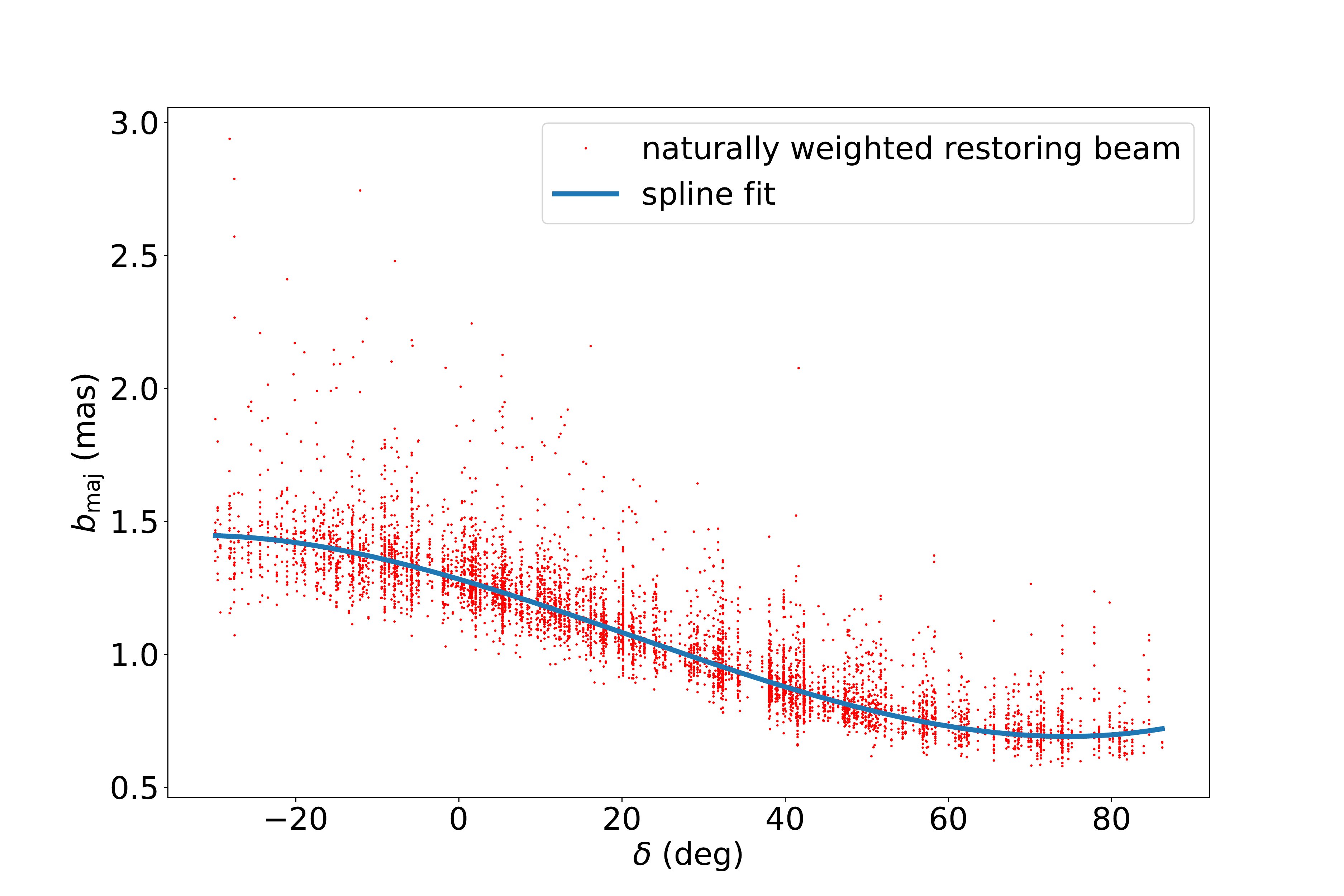}
    \includegraphics[width=\columnwidth]{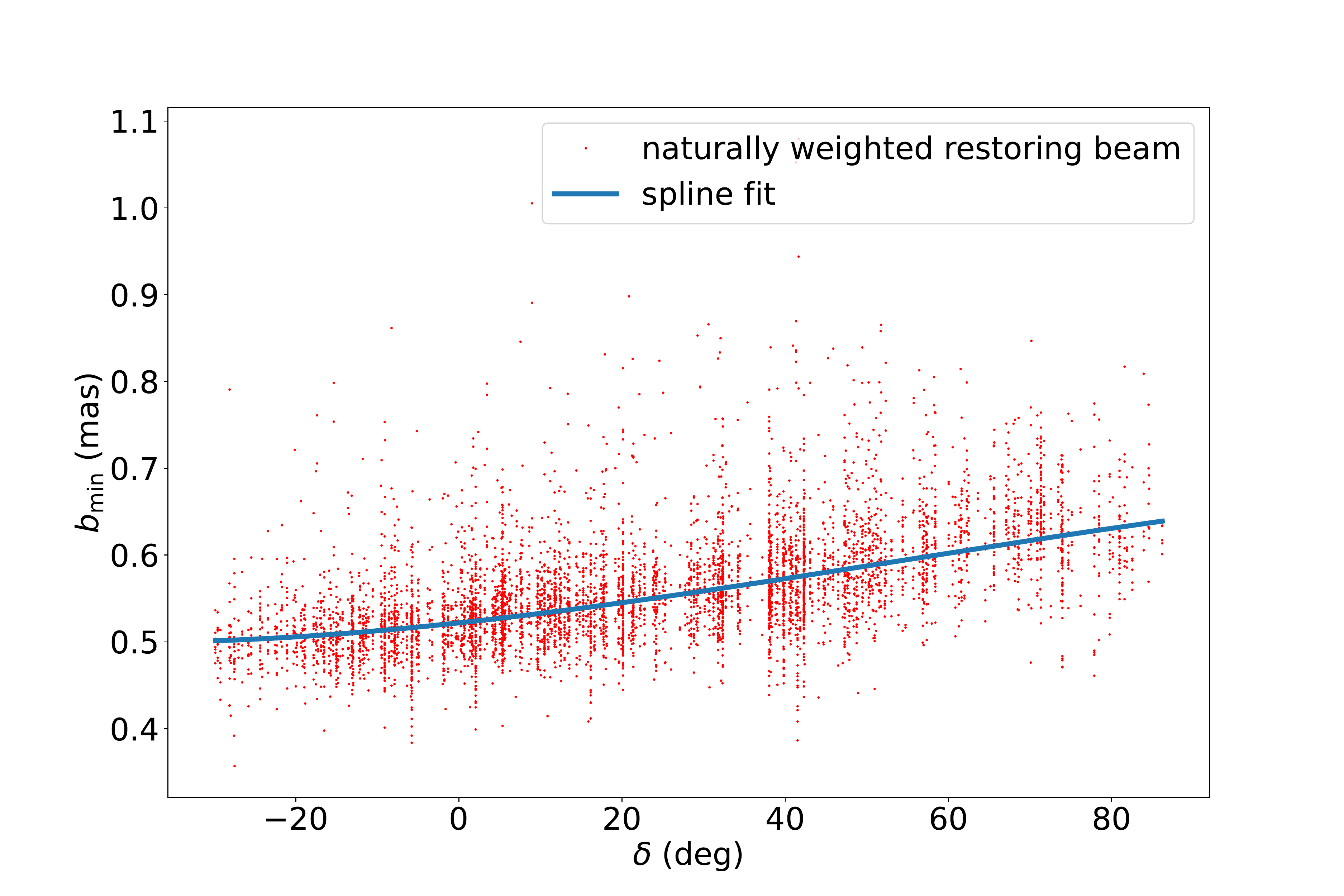}
    \caption{Spline fits for the maximum $b_\mathrm{maj}$ (top) and the minimum $b_\mathrm{min}$ (bottom) FWHM dimensions of the naturally weighted elliptical restoring beam vs declination $\delta$. Red points denote the restoring beam FWHM dimensions from the observations.
    \label{f:beam_fits}
    }
\end{figure}

We used spline fits to the major $b_\mathrm{maj}$ and minor $b_\mathrm{min}$ FWHM dimensions of the naturally weighted elliptical restoring beam with the source declination for several thousand 15~GHz VLBA observations made within the MOJAVE programme to define the size of the circular beam with which the $I$, $Q$ and $U$ maps are convolved. The fits for $b_\mathrm{maj}$ is 
\begin{equation}
1.28 - \delta\times8.95\times10^{-3} - \delta^2\times7.91\times10^{-5} + \delta^3\times1.24\times10^{-6}, 
\label{app_eq:bmaj}
\end{equation}
for  $b_\mathrm{min}$:
\begin{equation}
0.52 + \delta\times1.01\times10^{-3} + \delta^2\times8.88\times10^{-6} - \delta^3\times5.57\times10^{-8},  
\label{app_eq:bmin}
\end{equation}
where $\delta$ is the source J2000 declination in degrees. The plots of these functions together with the restoring beam FWHM dimensions from observations against declination are shown in Fig.~\ref{f:beam_fits}. Being a fixed array of ten antennas, the VLBA yields interferometric coverage which depends specifically on the source declination. The FWHM size of the circular Gaussian beam was taken as the arithmetic mean of $b_\mathrm{maj}$ and $b_\mathrm{min}$.

\section{Notes on individual AGN} \label{appendix:AGNs}

Here, we give the notes on AGNs which exhibit a significant anticorrelation  of median fractional polarization $m_{\mathrm{med}}$, a significant correlation of EVPA standard deviation $\sigma_{\mathrm{EVPA}}$ and a significant trend of relative fractional polarization variability $\sigma_m/m_{\mathrm{med}}$ with the distance from the core $r_{\mathrm{along}}$. The maps of polarization variability and $m_{\mathrm{med}}$ for the mentioned sources are given in the Supplementary data online.

0106+678: The significance of the positive correlations of the EVPA standard deviation and $\sigma_m/m_{\mathrm{med}}$ with $r_{\mathrm{along}}$ is driven by two pixels in the region at 3.5~mas from the core in which the polarization direction and the degree are less stable in comparison to the rest of the jet. The maps of individual epochs of the source show that the narrower the region with the detected polarization at 3.5~mas core separation, the better the EVPA aligns with the local jet direction. It might suggest that at different epochs the polarization was detected predominantly from the central spine or combined with polarized emission originated from the sheath surrounding the jet. The magnetic field pitch angle or viewing angle change can also contribute to the $\sigma_{\mathrm{EVPA}}$ and $\sigma_m/m_{\mathrm{med}}$ rise.

0118$-$272: Significant positive correlation of $\sigma_{\mathrm{EVPA}}$ vs $r_{\mathrm{along}}$ is due to non-uniformity on a scale comparable to the beam size and the uncertainty of the ridgeline construction near the jet edge. 

0316+413: A few pixels with large uncertainties cause a significant rise of $\sigma_m/m_{\mathrm{med}}$ along the ridgeline.

0415+379: Relative fractional polarization variability $\sigma_m/m_{\mathrm{med}}$ and EVPA standard deviation significantly drop along the ridgeline.

0430+052: The positive tendency of $\sigma_m/m_{\mathrm{med}}$ with the distance from the core appears to be intrinsic. The radio~galaxy also exhibits a significant increase of $m_\mathrm{med}$ with $r_{\mathrm{along}}$.

0738+313: The positive trend of relative fractional polarization variability down the jet is due to a bright feature on the averaged total intensity map. The EVPA standard deviation rises there as well.

0923+392: This quasar appears to have an intrinsic decrease of $\sigma_m/m_{\mathrm{med}}$. There is a region with both reduced  $\sigma_m/m_{\mathrm{med}}$ and $\sigma_{\mathrm{EVPA}}$ on one side of the bright feature on the $I_\mathrm{mean}$ map. This jet region manifests a transversely resolved complex structure and could be a mixture of several components or related to plasma turbulence.

1253$-$055: Median fractional polarization, EVPA standard deviation and $\sigma_m/m_{\mathrm{med}}$ decrease with $r_{\mathrm{along}}$ in this quasar.

1803+784: Relative fractional polarization variability rises with $r_{\mathrm{along}}$ because of non-uniformity on a beam scale.

1926+611: The significant positive trend of $\sigma_m/m_{\mathrm{med}}$ with $r_{\mathrm{along}}$ seems to be due to a jet bend.

2128$-$123: The source shows a significant anti-correlation of $m_{\mathrm{med}}$ with $r_{\mathrm{along}}$: $m_{\mathrm{med}}$ reduces and  $\sigma_{\mathrm{EVPA}}$ rises in the jet region around 6~mas.

2230+114: The positive significant correlation of $\sigma_{\mathrm{EVPA}}$ vs $r_{\mathrm{along}}$ is caused by the EVPA standard deviation increase in the quasi-stationary component at about 8~mas from the core (see fig.~3 for 2230+114 in \citet{Lister_2021}).

\bsp
\label{lastpage}
\end{document}